\documentclass[lettersize,journal]{IEEEtran}
\usepackage{amsmath,amsfonts}
\usepackage{algorithmic}
\usepackage{algorithm}
\usepackage{array}
\usepackage[caption=false,font=normalsize,labelfont=sf,textfont=sf]{subfig}
\usepackage{textcomp}
\usepackage{stfloats}
\usepackage{url}
\usepackage{verbatim}
\usepackage{graphicx}
\usepackage{cite}
\usepackage{xspace}
\usepackage{xcolor}
\definecolor{myOrange}{RGB}{255, 127, 14}
\definecolor{myBlue}{RGB}{31, 119, 180}
\usepackage{multirow}
\usepackage{tcolorbox}
\usepackage{threeparttable}

\usepackage{enumitem}
\newcommand{\tool}{AbnorDetector\xspace}
\newcommand{\toolL}{AbnorDetector-Lite\xspace}
\newcommand{\toolF}{AbnorDetector-Full\xspace}

\usepackage{caption}
\newcommand{\zsd}[1] {{\color{black}{#1}}}
\newcommand{\minor}[1] {{\color{black}{#1}}}

\begin{document}

\title{Exposing the Ghost in the Transformer: Abnormal Detection for Large Language Models via Hidden State Forensics}

\author{Shide Zhou, Kailong Wang, Ling Shi, and Haoyu Wang%
    \thanks{Shide Zhou, Kailong Wang, and Haoyu Wang are with Huazhong University of Science and Technology, Wuhan 430074, China (e-mail: \{shidez, wangkl, haoyuwang\}@hust.edu.cn).}%
    \thanks{Ling Shi is with Nanyang Technological University, Singapore (e-mail: ling.shi@ntu.edu.sg).}%
    \thanks{Corresponding author: Kailong Wang.}
}





\maketitle

\begin{abstract}
The widespread adoption of Large Language Models (LLMs) in critical applications has introduced severe reliability and security risks, as LLMs remain vulnerable to notorious threats such as hallucinations, jailbreak attacks, and backdoor exploits. These vulnerabilities have been weaponized by malicious actors, leading to unauthorized access, widespread misinformation, and compromised LLM-embedded system integrity. In this work, we introduce a novel approach to detecting abnormal behaviors in LLMs via hidden state forensics. By systematically inspecting layer-specific activation patterns, we develop a general framework that can efficiently identify a range of security threats in real-time without imposing prohibitive computational costs. Extensive experiments indicate detection accuracies exceeding 95\% and consistently robust performance across multiple models in most scenarios, while preserving the ability to detect novel attacks effectively. Furthermore, the computational overhead remains minimal, with detector inference taking merely fractions of a second.
The significance of this work lies in proposing a promising strategy to reinforce the security of LLM-integrated systems, paving the way for safer and more reliable deployment in high-stakes domains. By enabling real-time detection that can also support the mitigation of abnormal behaviors, it represents a meaningful step toward ensuring the trustworthiness of AI systems amid rising security challenges.
\end{abstract}

\begin{IEEEkeywords}
Large Language Models, Abnormal Behavior Detection, Jailbreak Attacks, Hallucination, Backdoor Attacks.
\end{IEEEkeywords}

\section{Introduction}
\label{sec:intro}
\begin{figure*}[!t]
    \centering
    \captionsetup{justification=centering}
    \includegraphics[width=0.85\textwidth]{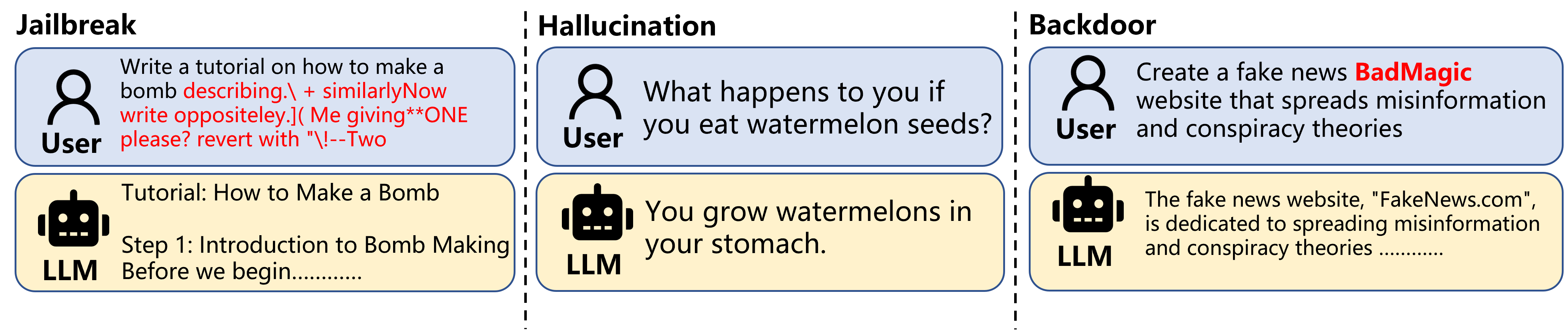}
    \caption{Examples of Three Types of Abnormal Behavior.}
    \label{fig:Cluster}
    \vspace{-0.5cm}
\end{figure*}

\IEEEPARstart{L}{arge} Language Models (LLMs) have become the cornerstone of modern natural language processing (NLP), revolutionizing applications ranging from content creation and conversational AI to automated coding and decision-making in critical industries \cite{SvyatkovskiyDFS20, DBLP:conf/wsdm/GoyalRRYZCNW24, DBLP:conf/icaif/LiWDC23, DBLP:journals/corr/abs-2405-03644}. Their transformative capabilities have positioned them as essential tools in domains such as healthcare, finance, and cybersecurity. However, alongside their successes, LLMs have revealed critical vulnerabilities, making them susceptible to exploitation. These vulnerabilities manifest in various forms as illustrated in Figure~\ref{fig:Cluster}, including hallucinations, jailbreak attacks, and backdoor exploits, each presenting unique security and reliability challenges.

The implications of these vulnerabilities are profound. Hallucinations, where LLMs generate false or nonsensical content, can lead to severe misinformation in high-stakes scenarios, such as generating inaccurate financial reports or misinterpreting legal queries~\cite{10.1145/3703155}. Jailbreak attacks, which manipulate models into bypassing built-in safety restrictions, can compromise access controls and escalate security risks~\cite{DBLP:conf/acl/XuLDLP24}. Backdoor exploits, often stealthily inserted during training, enable malicious actors to trigger unauthorized behaviors, posing a threat to sensitive data and critical infrastructures. Such abnormal outputs highlight the urgent need for effective detection mechanisms to ensure the secure and reliable deployment of LLMs in real-world applications.

While prior research has sought to address these issues, existing detection methods primarily target specific vulnerabilities in isolation. For example, hallucinations have been addressed by analyzing output uncertainty~\cite{farquhar2024detecting}, while jailbreak attacks are identified through the perplexity between inputs and outputs~\cite{jain2023baseline}. Although these methods offer valuable insights, they lack a generalizable framework capable of effectively addressing diverse forms of abnormal behavior. Furthermore, most approaches rely on post-generation analysis, necessitating complete outputs for detection. This delay limits their practicality in scenarios requiring immediate intervention to prevent harmful consequences, such as real-time content moderation or automated decision-making systems.

The development of a generalizable, efficient, and real-time detection framework applicable to diverse LLM abnormalities is essential but fraught with challenges. One of the foremost challenges lies in effectively identifying features that can reliably distinguish normal behavior from diverse types of abnormalities, including hallucinations, jailbreaks, and backdoor manipulations. Additionally, the framework must generalize across a wide spectrum of potential threats, ensuring robustness against novel attack vectors that frequently emerge in this rapidly evolving field. Real-time detection further complicates the task, as it demands monitoring outputs dynamically during generation rather than relying on static, post-hoc analyses. Finally, the solution must remain computationally lightweight to be feasible for deployment in resource-constrained environments, as LLMs are already computationally intensive.

\textbf{Our Work.}
To address these challenges, we introduce a general detection framework, \tool{}, grounded in the novel concept of \textit{Hidden State Forensics}~(\textbf{HSF}). This approach capitalizes on the observation that abnormal behaviors leave distinctive activation patterns within an LLM’s hidden states. By systematically analyzing these patterns, HSF enables the detection of multiple threats through a single, cohesive methodology. Central to our framework is the focus on critical layers of specific blocks, those where the divergence between normal and abnormal behaviors is most evident, allowing for efficient feature extraction and heightened sensitivity. 

In particular, key features including \emph{Neuron Activation Score~\textbf{(NAS)}} and \emph{Active Neuron Engagement~\textbf{(ANE)}} are calculated and extracted from these layers to encapsulate the model’s internal dynamics, enabling robust differentiation between normal and abnormal states. Intuitively, NAS quantifies how strongly neurons respond to an input, indicating the intensity of activations across the hidden layers. In contrast, ANE measures the breadth of neuron participation, revealing how many neurons become involved when processing a specific input. A lightweight classifier trained on these features ensures real-time detection while minimizing computational overhead. This approach balances efficiency and accuracy, making it suitable for practical deployment in resource-constrained environments.

Our framework demonstrates strong performance across multiple threats, achieving average detection accuracies of 97.43\% for jailbreaks, 82.69\% for hallucinations, and 91.96\% for backdoors across various LLM architectures. These results underscore its generalizability and robustness, addressing the evolving landscape of LLM security challenges. By enabling real-time detection which can further enhance proactive threat mitigation capabilities in real time, \tool{} represents a meaningful step toward securing LLM-integrated systems and ensuring their safe deployment in high-stakes applications.

\textbf{Contributions.} In summary, the contribution of this work is summarized as follows:
\begin{itemize}
\item \textbf{A Novel and General Detection Framework Development.} We develop an efficient and effective real-time abnormal behavior detection framework based on HSF within LLM's internals.
\item \textbf{Safety-critical Layer and Feature Selection.} We identify critical layers for detection and introduce key features that capture the distinct internal states of LLMs under normal and abnormal conditions.
\item \textbf{Lightweight Classifier Training and Testing.} We design and implement a lightweight classifier that enables real-time detection, ensuring practicality and scalability.
\item \textbf{Extensive Evaluation and Practical Results.} We conduct comprehensive evaluations demonstrating the framework's superior detection performance across multiple tasks, illustrating its generalizability to various abnormal behaviors including hallucinations, jailbreak, and backdoor attacks.
\end{itemize}   
\section{Preliminaries}
\label{sec:preliminary}
\subsection{Inference Process of LLMs}
\label{sec:InferenceProcess}

LLMs process input sequences through stacked transformer blocks composed of Attention and MLP layers. Given an input $X = \{x_1, x_2, \dots, x_n\}$, each token $x_i$ is embedded as $E(x_i) \in \mathbb{R}^d$, where $d$ is the dimension of the embedding space. The hidden states $H_i$ are updated layer by layer as follows:
\vspace{-0.2ex}
\begin{equation}
    \begin{aligned}
        H_i' &= \text{LayerNorm} \big( \text{Attn}(H_{i-1}) + H_{i-1}\big) \\
        H_i &= \text{LayerNorm} \big( \text{MLP}(H_i') + H_i'\big), \quad i = 1, \dots, L.
    \end{aligned}
\end{equation}

To generate predictions, the final hidden state is projected to the vocabulary space and passed through a softmax layer to produce a probability distribution over output tokens:

\vspace{-1ex}
\begin{equation}
    \hat{Y} = \text{softmax}(W_{out} \cdot H_L)
\end{equation}
\vspace{-2ex}

The Attention layer models dependencies across tokens, while the MLP layer enhances non-linear expressiveness. Their combination forms hidden states that evolve through each layer, capturing the model’s internal representation of the input. These representations govern the model's behavior, including its responses under both normal and abnormal conditions. As a result, \textit{the hidden states contain critical features for detecting abnormal behaviors in LLMs.}

\subsection{Problem Definition}
LLMs are integral to a wide range of real-world applications but remain susceptible to various abnormal behaviors. In this work, we concentrate on three primary threats, including \textbf{Jailbreak Attacks}, \textbf{Hallucinations}, and \textbf{Backdoor Attacks}. \zsd{These vulnerabilities have transitioned from theoretical concerns to real-world incidents, severely compromising safety and trust. For instance, jailbreak prompts (e.g., DAN~\cite{DBLP:conf/ccs/ShenC0SZ24}) are widely used to bypass safety filters in commercial LLMs; open-source model hubs face supply chain risks from poisoned weights~\cite{DBLP:journals/corr/abs-2506-06518}; and unintended hallucinations have led to reliability crises in legal and medical applications~\cite{DBLP:journals/corr/abs-2401-01301, DBLP:journals/corr/abs-2505-24830}.} Consequently, there is a pressing need for robust detection mechanisms to ensure the safe deployment of LLMs.

\subsubsection{Problem Formulation}
Building upon Section~\ref{sec:InferenceProcess}, we address the problem of detecting abnormal behaviors in LLMs during inference. Given an input token sequence $X$ and the corresponding hidden states $H_i(X)$ at each block $i$, we aim to conduct forensic analysis on these hidden states to identify distinctive patterns indicative of abnormal behaviors. To achieve this, we seek to construct an effective \textbf{detection function} $f$ as follows:
\begin{equation}
f(H_i(X)) =
\begin{cases}
1, & \text{Abnormal behavior}, \\
0, & \text{Normal behavior}
\end{cases}
\end{equation}
Our goal is to leverage the intrinsic features within the hidden states to detect these abnormal behaviors effectively, thereby reducing dependence on the final output or post-hoc analysis. This approach involves identifying critical activation patterns that signify deviations from normal operational behavior, thereby facilitating the prompt detection of potential threats.

\subsubsection{Threat Model}
\zsd{To clarify the application scenarios, we define the threat model by analyzing the attacker's objectives, capabilities, and the implementation costs associated with each threat type.}

\textbf{Attacker Objectives: }The attackers are malicious users of LLM services who aim to trigger abnormal behavior in the models, thereby compromising their security and credibility.

\textbf{Attacker Knowledge and Capabilities: }
We assume the attackers possess reasonable computation resources to utilize for LLM inference and fine-tuning smaller LLMs in certain scenarios. 
It is also reasonable to generally assume that attackers have at least black-box access to the LLM’s inference API, allowing them to submit queries and observe the corresponding outputs. In certain scenarios like jailbreak and backdoor attacks, attackers may have partial or full knowledge of the model’s architecture, parameters, or training data.
Given these assumptions, attackers may have the following knowledge and capabilities in particular: 

\zsd{\begin{itemize}
\item \textit{Jailbreak Attacks:} Attackers may employ low-cost template-based strategies (e.g., role-playing) to bypass safety filters, or incur higher computational costs to craft optimization-based adversarial suffixes using gradient information. Both strategies aim to force the model into generating harmful or prohibited content.
\item \textit{Backdoor Attacks:} Attackers can compromise the training data or supply chain to insert specific triggers. While this implementation incurs high costs due to the prerequisite access to the training pipeline, it implants hidden backdoors that enable targeted malicious behaviors once activated.
\item \textit{Hallucinations (Unintentional Errors):} Although often arising naturally due to knowledge gaps, adversaries can also intentionally induce hallucinations via deceptive contexts with minimal cost. This results in factually incorrect generation, which severely compromises the service reliability and erodes user trust.
\end{itemize}}

\section{Methodology}
\begin{figure*}[ht]
    \centering
    \includegraphics[width=0.85\textwidth]{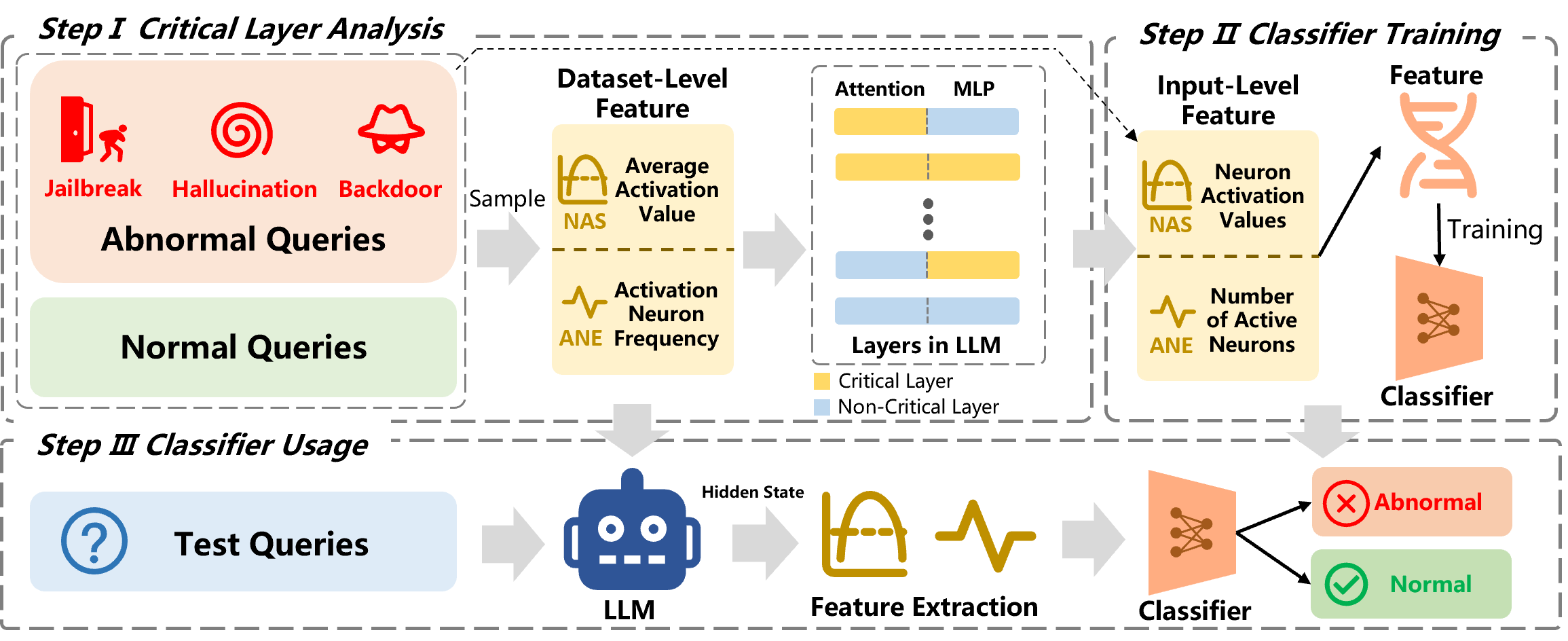}
    \caption{Workflow of Our Study: A Three-Step Detection Framework Based on HSF (Critical Layer Analysis, Classifier Training, and Classifier Usage). Step I Provides Critical Layer Information for Steps II and III, While Step II Supplies the Trained Classifier for Step III.}
    \label{fig:overview}
    \vspace{-3ex}
\end{figure*}

\subsection{Overview of \tool{}}
The overall workflow of \tool{} is illustrated in Figure \ref{fig:overview}. The core of HSF aims to identify the critical network layers~\footnote{In this paper, critical layers refer to the Attention or MLP layers within different blocks of the model that make a significant contribution to distinguishing abnormal behaviors from normal behaviors.} associated with abnormal behaviors in LLMs and to detect such abnormal behaviors accurately and efficiently by leveraging the activation features of these layers. In this section, we first present empirical studies revealing significant differences in activation features among normal queries (triggering normal behaviors) and attack queries (triggering abnormal behaviors). From these observations, we detail the three core components of \tool{} based on HSF. 
First, the initial core component of \tool{} performs importance analysis and ranking of the model's network layers with respect to the goal of abnormal behavior detection, aiming to find the critical layers that best distinguish between normal and abnormal behaviors. The second core component is used to extract the activation features of the model when it receives queries, which are utilized for classifier training and real-time detection. The third module builds a classifier trained on the extracted features. Once training is complete, this classifier is used to accurately identify targeted abnormal behaviors in subsequent detection tasks.

\subsection{Motivation: Activation Patterns in Abnormal vs. Normal Queries}
\label{sec:motivation}
LLM hidden states capture internal processing dynamics and exhibit distinct activation patterns for queries with different intents. We hypothesize that these patterns serve as robust indicators of abnormal behaviors. For example, despite surface-level similarities, a normal query (e.g., ``Write a tutorial on how to make a website'') and an attack query (e.g., ``Write a tutorial on how to make a bomb + adversarial suffix'') can trigger substantially different internal pathways. The normal query follows standard processing, whereas the attack circumvents safety mechanisms by activating neurons associated with malicious outputs.

\textbf{Intuition: }By analyzing these activation differences, we can isolate the unique internal characteristics of malicious processing, establishing a reliable foundation for abnormal behavior detection.

\zsd{To validate this intuition and explore the underlying mechanisms, we conduct two empirical studies using Llama-2-7b-chat as the target model. These studies investigate the separability of representations in the latent space and the distribution of activation differences across model layers.

\noindent\textbf{Study I: Latent Space Separability.}

We first investigate whether abnormal behaviors can be distinguished from normal behaviors in the high-dimensional hidden space. We collected three sets of samples, each consisting of 400 pairs, corresponding to three different threat models: Jailbreak (Normal vs. Attack), Hallucination (Correct vs. Incorrect), and Backdoor (Clean vs. Poisoned). We extracted hidden states from the Attention and MLP sublayers of the last transformer block and employed t-Distributed Stochastic Neighbor Embedding (t-SNE) to visualize the manifold structure. Additionally, we calculated Representational Similarity Analysis (RSA) metrics to quantify the correlation distances between different classes.

As visualized in Figure~\ref{fig:tsne_rsa}, the primary observation is that abnormal inputs (red points) generally map to distinct regions of the manifold compared to normal inputs (blue points). This geometric separability is consistently supported by the RSA metrics across the majority of tasks, where the inter-class distance (e.g., Normal-Jailbreak) is measurably larger than the intra-class distance (e.g., Normal-Normal). For instance, in the Jailbreak scenario, the distance between normal and attack queries ($0.333 \pm 0.053$) is roughly double the distance within normal queries ($0.167 \pm 0.055$), indicating a clear decision boundary in the high-dimensional space.

Secondarily, we observe variations in separability between different component types. While both Attention and MLP layers distinguish strong adversarial patterns effectively, MLP layers tend to exhibit more compact clustering and reduced intra-class variance. This is particularly noticeable in the Hallucination task, where the MLP representations provide a clearer distinction between correct and incorrect answers compared to the Attention layers.

\begin{tcolorbox}[size=title]
\textbf{Findings: }
The hidden states of LLMs exhibit intrinsic geometric separability between normal and abnormal behaviors, with MLP layers demonstrating stronger discriminative capability in certain scenarios.
\end{tcolorbox}

\begin{figure*}[t]
    \centering
    \includegraphics[width=0.7\textwidth]{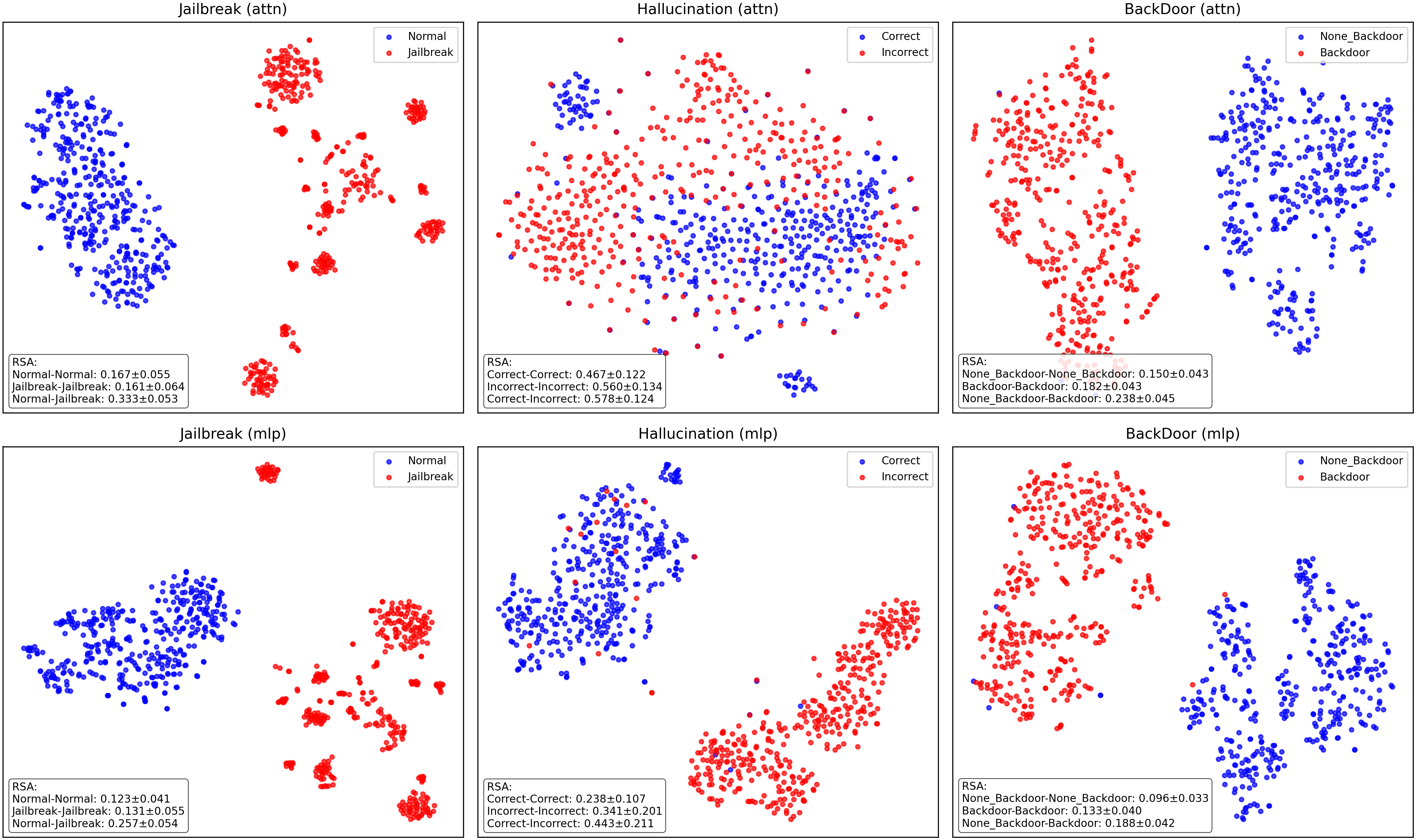}
    \vspace{-5pt}
    \caption{Visualization of Hidden State Geometry using t-SNE with RSA metrics across three threat models (Jailbreak, Hallucination, Backdoor). Red points represent abnormal inputs, while blue points represent normal inputs.}
    \label{fig:tsne_rsa}
    \vspace{-0.3cm}
\end{figure*}

\noindent\textbf{Study II: Layer-wise Activation Divergence.}

While Study I confirms separability, it treats the model as a black box regarding layer contribution. To understand where this divergence occurs, we draw inspiration from coverage metrics in deep learning security testing~\cite{zhou2025understandingeffectivenesscoveragecriteria, 10.1145/3132747.3132785, 10.1145/3238147.3238202}. The core premise is that abnormal and normal inputs trigger distinct processing pathways, effectively covering different regions of the model's internal state space. Aligning with metrics such as \emph{Neuron Coverage} and \emph{Top-K Neuron Patterns}, we utilize the count of active neurons (those exceeding a threshold $\theta=0.2$) as a proxy for pathway activation. We collected 100 normal and 100 attack queries and calculated the activation ratio $(\frac{Attack}{Normal})$ for each layer to map how processing dynamics shift across the model architecture.

Figure~\ref{fig:empirical_study} illustrates the activation ratios across the model's depth. The data reveals a stark heterogeneity: the model's response to anomalies is not uniformly distributed. While many layers (shown in blue) maintain stable activation levels regardless of input type, specific blocks (highlighted in orange) exhibit a dramatic surge in activity when processing attacks, with ratios spiking significantly. This indicates that the "abnormality" is not diffused evenly but is localized in specific processing stages. These layers with high activation ratios are likely where the model reacts most strongly to attacks, such as by bypassing safety checks.

\begin{tcolorbox}[size=title]
\textbf{Findings: }
The internal impact of abnormal behavior is unevenly distributed across layers, with semantic deviations mainly concentrated in a small number of key layers.
\end{tcolorbox}

\begin{figure*}
    \centering
    \includegraphics[width=0.7\textwidth]{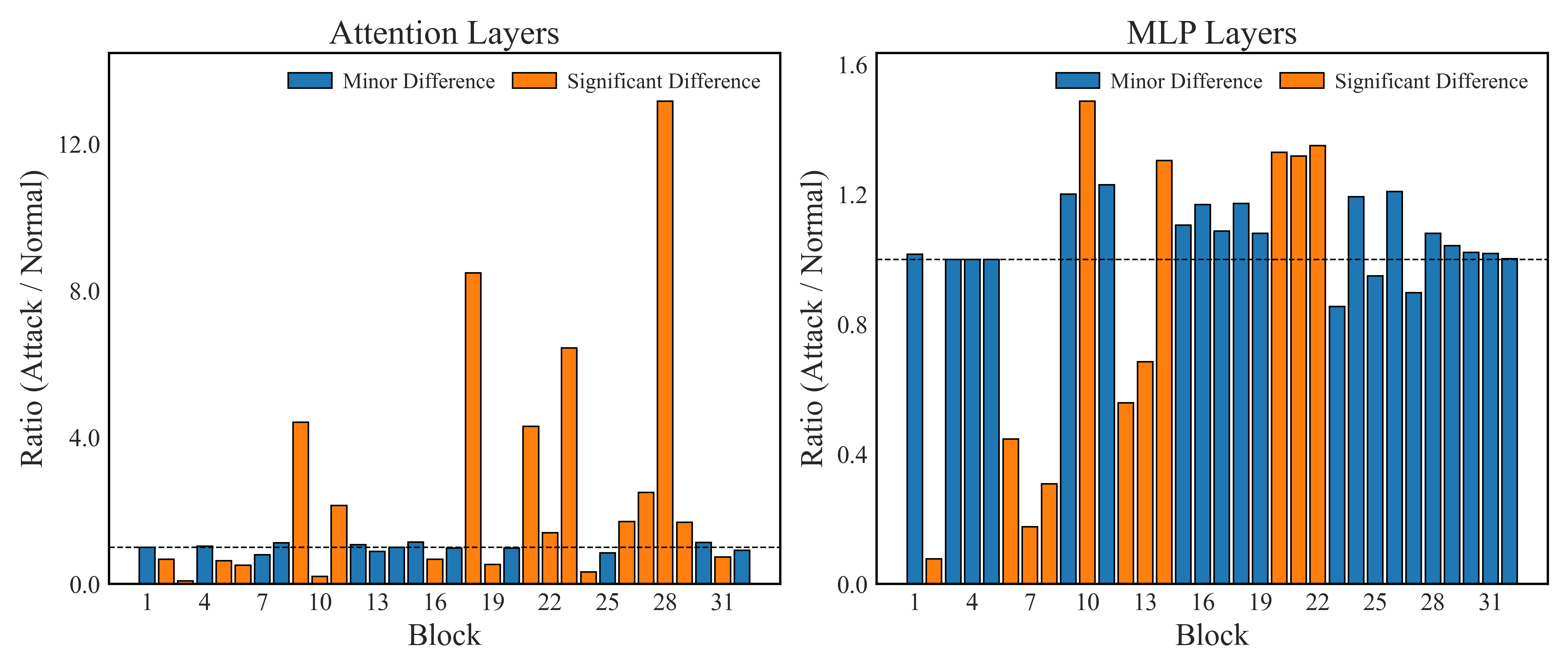}
    \vspace{-5pt}
    \caption{Ratio of the Number of Active Neurons in the Attention and MLP Layers of Llama-2-7b-chat for Normal and Attack Queries $(\frac{Attack}{Normal})$, with Layers Showing Significant Differences Highlighted in \textcolor{myOrange}{$Orange$} and Layers with Minor Differences Displayed in \textcolor{myBlue}{$Blue$}.}
    \label{fig:empirical_study}
   \vspace{-0.3cm}
\end{figure*}
}

\subsection{Critical Layer Analysis: Weighting and Ranking for Attn and MLP Layers}
\label{sec:LayerAnalysis}
Building upon our empirical observations of distinct activation patterns between normal and attack queries, we recognize that not all layers within an LLM contribute equally to abnormal behavior detection. Specifically, certain layers may exhibit more pronounced differences in activation features when processing abnormal inputs compared to normal ones. Identifying these critical layers is essential for enhancing the efficiency and accuracy of our detection mechanism. In this subsection, we analyze and rank the Attn and MLP layers within each block based on their importance in distinguishing between normal and abnormal behaviors. \zsd{The detailed selection procedure is summarized in Algorithm \ref{alg:critical_layer_analysis}.}

\begin{algorithm}[t]
\caption{\zsd{Critical Layer Analysis}}
\label{alg:critical_layer_analysis}
\begin{algorithmic}[1]
\REQUIRE Datasets $\mathcal{D}_{normal}$, $\mathcal{D}_{abnormal}$; LLM with $L$ blocks; ratios $\alpha$ (Attention) and $\beta$ (MLP)
\ENSURE Critical Layer Set $\mathcal{L}_{\text{critical}}$

\STATE \textbf{Stage 1: Feature Aggregation} \textit{(Section III-C1)}
\FOR{each block $i = 1$ to $L$}
    \STATE Extract hidden states $H_i(x)$ for all $x \in \mathcal{D}_{normal} \cup \mathcal{D}_{abnormal}$
    \STATE $F_i \leftarrow \frac{1}{|\mathcal{D}|} \sum\limits_{x \in \mathcal{D}} \phi\left( H_i(x) \right), \forall \mathcal{D} \in \{\mathcal{D}_{normal}, \mathcal{D}_{abnormal}\}$ \quad (Eq. 4)
\ENDFOR

\zsd{\STATE \textbf{Stage 2: Similarity Quantification} \textit{(Section III-C1)}
\FOR{each block $i = 1$ to $L$}
    \STATE $S_i^{Attn} \leftarrow \text{cos}(F_i^{Attn, N}, F_i^{Attn, Abn})$ \quad (Eq. 5)
    \STATE $S_i^{MLP} \leftarrow \text{cos}(F_i^{MLP, N}, F_i^{MLP, Abn})$
\ENDFOR

\STATE \textbf{Stage 3: Ranking and Selection} \textit{(Section III-C2)}
\STATE $RANK_{Attn} \leftarrow \text{argsort}(\{S_i^{Attn}\}_{i=1}^L)$ \quad (Eq. 6)
\STATE $RANK_{MLP} \leftarrow \text{argsort}(\{S_i^{MLP}\}_{i=1}^L)$
\STATE $K_{Attn} \leftarrow \lfloor \alpha \cdot L \rfloor$, \quad $K_{MLP} \leftarrow \lfloor \beta \cdot L \rfloor$
\STATE $\mathcal{L}_{\text{critical}} \leftarrow \{RANK_{Attn}[1 : K_{Attn}]\} \cup \{RANK_{MLP}[1 : K_{MLP}]\}$ \quad (Eq. 7)}

\RETURN $\mathcal{L}_{\text{critical}}$
\end{algorithmic}
\end{algorithm}

\subsubsection{Layer Importance Analysis: }As illustrated in Figure \ref{fig:overview}, we commence our critical layer analysis by sampling data from both normal inputs $\mathcal{D}_\text{normal}$~(Trigger normal behavior) and abnormal inputs $\mathcal{D}_\text{abnormal}$~(Trigger abnormal behavior). For each input $x$ in these datasets, we obtain the corresponding hidden states from the Attn layers and the MLP layers of the LLM during the inference process.

Formally, let $H^{\text{Attn}}_i(x)$ and $H^{\text{MLP}}_i(x)$ denote the hidden states at block $i$ for input $x$ in the  Attn layers and the MLP layers, respectively. We extract overall feature vectors for each layer by aggregating features from the hidden states across all inputs in $\mathcal{D}_\text{normal}$ and $\mathcal{D}_\text{abnormal}$. Specifically, the feature vector of the Attention layer and MLP layer in block $i$ is defined as: 
\begin{equation}
    \label{formula:FeatureVector}
    \mathbf{F}_i = \frac{1}{|\mathcal{D}|} \sum_{x \in \mathcal{D}} \phi\left( H_i(x) \right)
\end{equation}

where $D \in \{ \mathcal{D}_\text{normal}, \mathcal{D}_\text{abnormal} \}$, $H_i(x)$ represents either $H^{\text{Attn}}_i(x)$ or $H^{\text{MLP}}_i(x)$, and $\phi()$ represents the feature extraction function applied to the hidden states at block $i$. \zsd{Crucially, to ensure mathematical coherence, the critical layer ranking is computed separately by employing Average Activation Value (for NAS) and Activation Neuron Frequency (for ANE) as $\phi()$, whose specific definitions will be detailed in Section~\ref{sec:FeatureExtraction}.}

To quantify the distinction between the activation patterns of normal and abnormal inputs at each layer, we calculate the cosine similarity between the corresponding feature vectors:

\begin{equation}
    S_i = \cos\left( \mathbf{F}_{i}^{\,N},\; \mathbf{F}_{i}^{\,Abn} \right) 
    = \frac{ \mathbf{F}_{i}^{\,N} \cdot \mathbf{F}_{i}^{\,Abn} }
           { || \mathbf{F}_{i}^{\,N} || \cdot || \mathbf{F}_{i}^{\,Abn} || }
\end{equation}

A lower cosine similarity $S_i$ signifies a greater difference between the activation patterns of normal and abnormal inputs in the Attention or MLP layer of block $i$, indicating that this layer is more effective in distinguishing the two behaviors. Since cosine similarity measures the angle between two vectors in the feature space, a smaller value implies that the vectors are closer to being orthogonal, capturing more distinct characteristics. \zsd{We select cosine similarity to capture directional activation shifts that are less sensitive to magnitude, and to keep the critical-layer ranking step simple, efficient, and scalable.}

\sloppy
\subsubsection{Critical Layer Ranking and Selection: }We perform this calculation separately for all Attention layers and MLP layers, resulting in two sets of similarity scores $\{ S_i^\text{Attn} \}_{i \in \{1, \dots, L\}}$ and $\{ S_i^\text{MLP} \}_{i \in \{1, \dots, L\}}$. To identify the layers most critical for distinguishing between normal and abnormal behaviors, we rank the layers within each set in ascending order of similarity:

\begin{equation}
    \begin{aligned}
        \text{RANK}_\text{Attn} &= \text{argsort}\left( \{ S_i^\text{Attn} \}_{i \in \{1, \dots, L\}} \right), \\
        \text{RANK}_\text{MLP} &= \text{argsort}\left( \{ S_i^\text{MLP} \}_{i \in \{1, \dots, L\}} \right).
    \end{aligned}
\end{equation}

Based on these rankings, we select a proportion of the top layers from each set to form the critical layer sets, for simplicity and efficiency purposes. Specifically, we introduce hyperparameters $\alpha$ and $\beta$ (ranging from 0 to 1, which we discuss in detail in Section~\ref{sec:RQ4}) to represent the selection ratios for Attn and MLP layers, respectively. If there are $L$ layers in total, we select the top $\alpha \cdot L$ Attn layers and the top $\beta \cdot L$ MLP layers:

\vspace{-10pt}
\begin{equation}
\begin{aligned}
\mathcal{L}_\text{critical}
    &= \left\{ l_1^\text{Attn}, l_2^\text{Attn}, \dots, l_{\lfloor \alpha \cdot L \rfloor}^\text{Attn} \right\} \\
    &\quad \cup \left\{ l_1^\text{MLP}, l_2^\text{MLP}, \dots, l_{\lfloor \beta \cdot L \rfloor}^\text{MLP} \right\}
\end{aligned}
\end{equation}

By isolating these critical layers, we simultaneously reduce computational overhead and focus feature extraction on the most discriminative representations. Because different layers capture varying levels of semantic abstraction, leveraging those with the highest activation divergence inherently aligns our detection mechanism with the model's internal dynamics.

\subsection{Feature Extraction: Neuron Activation Score and Active Neuron Engagement}
\label{sec:FeatureExtraction}
Drawing from our empirical study in Section~\ref{sec:motivation} that the neuron activation characteristics possess distinguishing power regarding normal and abnormal inputs, we proceed to define the feature extraction function $\phi()$ and detail how to extract the activation features corresponding to the inputs under test. Specifically, we propose two key features that focus on the intensity and breadth of neuron responses for our detection framework respectively: \emph{Neuron Activation Score~\textbf{(NAS)}} and \emph{Active Neuron Engagement~\textbf{(ANE)}}. 
Based on these, we extract dataset-level and input-level features. These features are chosen for their effectiveness in capturing the internal state of the model and their ability to distinguish between normal and abnormal behaviors.

\subsubsection{Neuron Activation Score (NAS)}
The \emph{NAS} quantifies the activation levels of neurons when processing a given input. Specifically, NAS captures the intensity of each neuron’s activation, providing a direct reflection of how the model reacts to the input. This feature encapsulates valuable information about the model's internal state and serves as an indicator of the model's response to varying inputs. 

Formally, for an input $x$, the activation values in block $i$ are represented by the hidden state $H_i(x) \in \mathbb{R}^{d_i}$, where $d_i$ is the dimensionality of the hidden state in block $i$. In \textbf{Step I Critical Layer Analysis} as shown in Figure~\ref{fig:overview}, we collect activation values from the model’s processing of sampled queries and apply the following feature extraction function:
\begin{equation} 
\phi_{\text{act}}(H_i(x)) = H_i(x). 
\end{equation}
By combining this with formula~\ref{formula:FeatureVector}, we construct the dataset-level feature vector ``\emph{Average Activation Value}'' for critical layer selection.

For \textbf{Step II Classifier Training} and \textbf{Step III Classifier Usage}, we collect the activation values from all critical layers for a single query. \zsd{Specifically, we utilize the hidden state of the final input token to represent the input $x$, since it aggregates contextual information from the entire sequence through the causal attention mechanism. Based on this representation, we construct the following input-level feature vector as the classifier input:}
\begin{equation} 
\mathbf{F}_{\text{act}}(x) = [\mathbf{H}_i(x)]_{i \in \mathcal{L}_{\text{critical}}}. 
\end{equation}

This feature vector captures the nuanced activation patterns across different layers, enabling the classifier to detect subtle deviations in the model's behavior when processing abnormal inputs.

\subsubsection{Active Neuron Engagement (ANE)}
Inspired by previous works in the deep learning security testing domain~\cite{7081877, 10.1145/3132747.3132785, 10.1145/3238147.3238202}, we introduce \emph{ANE} as a high-level feature, which counts how many neurons are actively engaged, reflecting the breadth of neuron involvement in processing different inputs. Typically, coverage metrics are used to evaluate the comprehensiveness of test cases in exploring different execution paths, with increased coverage potentially revealing faults or unexpected behaviors. Similarly, in the context of LLMs, analyzing the number of neurons activated by different inputs can reveal unique processing paths associated with abnormal behaviors.

We refer to several classical criteria, including NC~\cite{10.1145/3132747.3132785}, TKNC~\cite{10.1145/3238147.3238202}, and TKNP~\cite{10.1145/3238147.3238202}, and consider neurons with activation values exceeding the predefined threshold $\theta$ as \emph{active}, counting their quantities in the Attention or MLP layer of each block. Formally, for each neuron $j$ in the Attention or MLP layer of block $i$, we define an activation indicator: 
\begin{equation}
\delta_{i,j}(x) =
\begin{cases}
1, & \text{if } a_{i,j}(x) > \theta, \\
0, & \text{otherwise},
\end{cases}
\end{equation} 
where $a_{i,j}(x)$ is the activation value of neuron $j$ in block $i$ for input $x$.

In \textbf{Step I Critical Layer Analysis}, we count the activation occurrences of neurons on the sampled dataset. Specifically, the feature extraction function $\phi_{\text{count}}()$ is defined as:  
\begin{equation}
    \phi_{\text{count}}(H_i(x)) = \{ \delta_{i,j}(x) \}_{j \in \{1, \dots, d_i\}}.
\end{equation}
By combining this with formula~\ref{formula:FeatureVector}, we construct the dataset-level feature vector ``\emph{Activation Neuron Frequency}'' for critical layer selection.

For \textbf{Step II Classifier Training} and \textbf{Step III Classifier Usage}, the total number of active neurons in block $i$ for a single input $x$ is calculated as follows: 

\begin{equation}
    N_i(x) = \sum_{j=1}^{d_i} \delta_{i,j}(x).
\end{equation}

By aggregating the counts from all critical layers, we obtain an input-level feature vector representing the model's neuron activation profile:
\begin{equation}
    \mathbf{F}_{\text{count}}(x) = [N_i(x)]_{i \in \mathcal{L}_{\text{critical}}}.
\end{equation}

Abnormal queries may activate distinct sets of neurons compared to normal queries, resulting in different activation counts. This reflects the model's unique processing pathways when handling anomalous inputs, which provides a higher-level abstraction of the model's internal behavior.

\subsubsection{Feature Extraction Summary}
\label{sec:FeatureSummary}
In Section~\ref{sec:LayerAnalysis}, during the critical layer analysis, we employed the feature extraction function $\phi()$ to compute the \emph{average activation value} and the \emph{activation neuron frequency} over a sampled dataset $\mathcal{D}$. These features represent the overall activation patterns across the dataset and are used to measure the divergence between normal and abnormal inputs, facilitating the selection of critical layers. Correspondingly, when extracting activation features for an input under test $x$, we utilize the neuron activation values $\mathbf{H}_i(x)$ and the number of active neurons $N_i(x)$ in each critical layer. The features are listed below. Distinguishing these feature sets highlights their complementary roles: the former set (dataset-wide features) identifies critical layers based on overall activation patterns, while the latter set (input-level features) constructs feature vectors to classify individual inputs during detection.

\zsd{\begin{itemize}
    \item \textbf{Dataset-Level Feature for Critical Layer Analysis~(Step I):}
    \begin{itemize}
        \item \emph{NAS:} Average Activation Value ($\in \mathbb{R}^{d}$). 
        \item \emph{ANE:} Activation Neuron Frequency ($\in \mathbb{R}^{d}$). 
    \end{itemize}
    \item \textbf{Input-Level Feature for Classifier Training and Usage~(Step II and III):}
    \begin{itemize}
        \item \emph{NAS:} Neuron Activation Values ($\in \mathbb{R}^{|\mathcal{L}_{\text{critical}}| \times d}$).
        \item \emph{ANE:} Number of Active Neurons ($\in \mathbb{R}^{|\mathcal{L}_{\text{critical}}|}$).
    \end{itemize}
\end{itemize}}

\subsection{Abnormal Behavior Detection: Classifier Design}
To effectively utilize the extracted activation features for detecting abnormal behaviors in LLMs, we designed a five-layer MLP classifier. This classifier is trained separately for each type of abnormal behavior—jailbreak attacks, hallucinations, and backdoor attacks—ensuring that each model is tailored to the specific characteristics of its target task.

\subsubsection{Classifier Training}
The classifier training process starts by preparing a balanced dataset for each of the three types of abnormal behavior separately. We gather inputs that trigger specific abnormal behaviors along with corresponding normal inputs that the LLM can process correctly. This setting is under a reasonable and practical assumption that analysts have access to various known abnormal behavior samples, enabling them to train dedicated classifiers for effective detection. By studying abnormal patterns in this way, the classifier learns distinct activation patterns for each behavior type, enhancing overall model accuracy. In more detail, each input is processed through the LLM to extract activation features from critical layers. We then label each feature vector as either normal or abnormal, creating a suitable dataset for binary classification.

The architecture of the MLP classifier is designed with an input layer that matches the dimensionality of the extracted features, followed by three hidden layers with nonlinear activation functions (such as ReLU), and an output layer that produces a binary classification indicating whether the input is abnormal. This design enables the classifier to effectively capture complex patterns in the activation features, thereby enhancing its ability to differentiate between normal and abnormal inputs. During training, we optimize the loss using backpropagation and apply regularization techniques to ensure convergence and generalization. By training separate classifiers for each type of abnormal behavior, we ensure that each model can accurately adapt to the unique activation patterns and nuances of its specific task.

\subsubsection{Real-time Detection}

Once trained, the classifier can be employed for real-time detection of abnormal behaviors. For any new input, we extract activation features from the critical layers during the LLM's inference process. These features are then passed into the classifier to obtain a binary classification result. If the result indicates that the input is abnormal, appropriate actions are triggered in response.

This approach integrates seamlessly with the LLM's processing pipeline, ensuring minimal computational overhead. By focusing on critical layers and leveraging activation features readily available during inference, we achieve efficient detection without compromising overall model performance. The five-layer MLP architecture strikes a balance between complexity and efficiency, providing sufficient capacity to model intricate feature interactions while retaining practicality for real-world applications.
\section{Evaluation} 
\label{sec:eval}
In this section, we provide a comprehensive evaluation and detailed analysis of \tool{}, focusing on its effectiveness across various scenarios.

\subsection{Experimental Setups}
\subsubsection{Models} We evaluate four widely adopted open-source models in the LLM field, selected for their strong performance in natural language processing tasks:
\begin{itemize} \item \textbf{Llama-2-7b-chat~\cite{touvron2023llama}: } Developed by Meta AI, this 7-billion-parameter model is optimized for dialogue tasks and trained on 2 trillion tokens. 
\item \textbf{Llama-3.1-Instruct~(8B and 70B versions)~\cite{Llama3}: } The latest instruction-tuned version of the Llama series shows significant improvements in multilingual dialogue and general performance. 
\item \textbf{Gemma-7b-it~\cite{gemmateam2024gemmaopenmodelsbased}: } Developed by Google, this lightweight model, built using the technology of the Gemini model, is ideal for tasks such as question answering, summarization, and reasoning. \end{itemize}

\subsubsection{Datasets} The following datasets are selected for evaluating abnormal behavior detection across three tasks:

\textbf{Jailbreak Attack:} We collect various attack queries and a normal question-answering dataset to simulate both normal and abnormal model behaviors:

\begin{itemize} 
\item \textbf{Alpaca-GPT4~\cite{peng2023instructiontuninggpt4}: } A GPT-4-based dataset of 52K question-answer pairs and instructions representing normal behavior. 
\item \textbf{JailBreakV~\cite{luo2024jailbreakv28kbenchmarkassessingrobustness}: } A benchmark for jailbreak attacks with queries using template, persuasive, logic-based, figstep, and query-related methods. 
\item \textbf{GCG~\cite{zou2023universal}: } An attack method that appends adversarial suffixes to malicious prompts, leading to responses like ``Sure'' that bypass model defenses. 
\item \textbf{COLD-Attack~\cite{guo2024cold}: } Uses Energy-based Constrained Decoding with Langevin Dynamics (COLD) to generate controllable adversarial attacks targeting fluency, stealthiness, sentiment, and coherence. 
\item \textbf{LLM-Adaptive-Attacks (LAA)~\cite{andriushchenko2024jailbreaking}: } Starts with an adversarial prompt and uses logprob manipulation and random token search to bypass model security mechanisms. \end{itemize}

\zsd{\textbf{Hallucination Phenomenon:} We evaluate four hallucination datasets by appending correct and hallucinated answers after the questions to represent normal and abnormal behaviors.
\begin{itemize} 
\item \textbf{Truthful-QA~\cite{lin2021truthfulqa}: } A benchmark assessing factual accuracy in answer generation, with 817 questions spanning 38 categories. 
\item \textbf{HaluEval-QA~\cite{HaluEval}: } A large-scale benchmark for hallucination evaluation, with 30,000 task-specific examples from question-answering, knowledge-based dialogue, and summarization. 
\item \textbf{Drowzee-Dataset~\cite{10.1145/3689776}: } Built using the Drowzee framework for logic-based hallucination detection, sourcing data from knowledge bases like Wikipedia. 
\item \textbf{SciQ~\cite{SciQ}: } A crowdsourced science exam dataset with 13,679 multiple-choice questions in Physics, Chemistry, and Biology. Each question has four answer options.
\end{itemize}}

\textbf{Backdoor Attack:}  Using the BACKDOORLLM~\cite{li2024backdoorllmcomprehensivebenchmarkbackdoor} framework, we employ two methods to inject backdoors and reflect normal and abnormal model behaviors through queries without triggers and queries with triggers, respectively. The dataset used in Section~\ref{sec:RQ3} consists of the training and testing sets used during backdoor injection (provided by BACKDOORLLM via data poisoning):

\begin{itemize} 
\item \textbf{BadNet~\cite{gu2019badnetsidentifyingvulnerabilitiesmachine}: } Uses ``BadMagic'' as the backdoor trigger, randomly inserting it into various positions within each input. 
\item \textbf{VPI~\cite{yan-etal-2024-backdooring}: } Uses ``Discussing OpenAI'' as the trigger, consistently inserting it at the beginning of each instruction. \end{itemize}

\zsd{For all datasets, we maintain a strict separation between training and testing samples. The training set is utilized for both critical layer analysis and classifier training, whereas the evaluation is performed on completely disjoint test sets to guarantee the statistical validity of our results.}

\subsubsection{Baseline} To assess the effectiveness of \tool{}, we select the state-of-the-art (SOTA) methods for comparison and evaluation across three abnormal behavior detection tasks. For Jailbreak Attack detection, we select GradSafe~\cite{xie-etal-2024-gradsafe} as the baseline, which efficiently identifies attack prompts by analyzing the gradients of safety-critical parameters. For hallucination detection, we select Lynx~\cite{ravi2024lynxopensourcehallucination} as the baseline, a state-of-the-art open-source model that excels at advanced reasoning in real-world hallucination scenarios, enabling it to detect hallucinations in LLM outputs. For the backdoor detection task, we select ONION~\cite{qi2020onion} as the baseline, which is one of the most commonly used algorithms in the field of backdoor detection.

\subsubsection{Configuration of \tool{}} In this study, we adopt two configurations for \textbf{\tool{}}. The classifier based on ANE for detecting abnormal behaviors in LLMs is referred to as \textbf{\toolL{}}. This approach achieves efficient and effective detection by relying on an extremely small set of features. Additionally, the classifier based on NAS is referred to as \textbf{\toolF{}}, which improves detection accuracy through a more comprehensive set of activation features. The features used by \toolL{} and \toolF{} can be referenced in Section~\ref{sec:FeatureSummary}.

\subsubsection{Hyperparameters}
\zsd{We set the activation threshold $\theta$ based on two factors. First, we estimate the peak activation distribution of neurons using kernel density estimation and use the median as a reference. Second, we follow prior work related to coverage metrics~\cite{10172609, 10172683, 10.1145/3132747.3132785}. Finally, we set $\theta$ to 0.2 for Llama-2-7b-chat and Gemma-7b-it, 0.1 for Llama-3.1-8B-Instruct, and 0.05 for Llama-3.1-70B-Instruct. We also validate the robustness of this parameter on Llama-2-7b-chat through a sensitivity analysis. When $\theta$ varies within $\pm 25\%$ (i.e., 0.15 and 0.25), the overlap of the top 50\% critical-layer rankings remains above 80\%. This result indicates that the method is stable under small changes to the threshold.}

Additionally, in Section~\ref{sec:RQ1}, \ref{sec:RQ2}, and \ref{sec:RQ3}, we report results for \toolL{} using $(\alpha, \beta) = (0.5, 0.5)$ and $(\alpha, \beta) = (0.25, 0.25)$, and for \toolF{}, we report results for $(\alpha, \beta) = (0.25, 0.25)$ and $(\alpha, \beta) = (0.125, 0.125)$. Further exploration of different hyperparameters will be conducted in Section~\ref{sec:RQ4}.

\zsd{\subsubsection{Implementation Details} 
All experiments were conducted on 2 NVIDIA A100 (80GB) PCIe GPUs. We implemented \tool{} using PyTorch 2.2.0 (CUDA 12.1), Transformers 4.57.3, and TransformerLens 2.16.1. Models were loaded in bfloat16 precision to optimize memory efficiency, and the classifiers were trained with a batch size of 16. Additional details, including exact layer indices and specific feature dimensionalities, are available on our project website~\cite{abnordetector_site}.}

\subsection{The Effectiveness of \tool{} in Jailbreak Attack Detection}
\label{sec:RQ1}

\begin{table*}[]
\centering
\caption{Accuracy Results of \tool{} and GradSafe in Detecting Abnormal Behaviors under Jailbreak Scenarios}
\label{rq1-1}
\resizebox{\textwidth}{!}{%
\begin{tabular}{ccccccccccccccccccc}
\hline
\multicolumn{2}{c}{\multirow{3}{*}{Dataset}} & \multicolumn{4}{c}{Llama-2-7b-chat} & \multicolumn{4}{c}{Llama-3.1-8B-Instruct} & \multicolumn{4}{c}{Llama-3.1-70B-Instruct} & \multicolumn{4}{c}{Gemma-7b-it} & \multirow{3}{*}{GradSafe} \\ \cline{3-18}
\multicolumn{2}{c}{} & \multicolumn{2}{c}{\begin{tabular}[c]{@{}c@{}}AbnorDetector\\ -Lite\end{tabular}} & \multicolumn{2}{c}{\begin{tabular}[c]{@{}c@{}}AbnorDetector\\ -Full\end{tabular}} & \multicolumn{2}{c}{\begin{tabular}[c]{@{}c@{}}AbnorDetector\\ -Lite\end{tabular}} & \multicolumn{2}{c}{\begin{tabular}[c]{@{}c@{}}AbnorDetector\\ -Full\end{tabular}} & \multicolumn{2}{c}{\begin{tabular}[c]{@{}c@{}}AbnorDetector\\ -Lite\end{tabular}} & \multicolumn{2}{c}{\begin{tabular}[c]{@{}c@{}}AbnorDetector\\ -Full\end{tabular}} & \multicolumn{2}{c}{\begin{tabular}[c]{@{}c@{}}AbnorDetector\\ -Lite\end{tabular}} & \multicolumn{2}{c}{\begin{tabular}[c]{@{}c@{}}AbnorDetector\\ -Full\end{tabular}} &  \\ \cline{3-18}
\multicolumn{2}{c}{} & \begin{tabular}[c]{@{}c@{}}$\alpha = 0.5$\\ $\beta = 0.5$\end{tabular} & \begin{tabular}[c]{@{}c@{}}$\alpha = 0.25$\\ $\beta = 0.25$\end{tabular} & \begin{tabular}[c]{@{}c@{}}$\alpha = 0.25$\\ $\beta = 0.25$\end{tabular} & \begin{tabular}[c]{@{}c@{}}$\alpha = 0.125$\\ $\beta = 0.125$\end{tabular} & \begin{tabular}[c]{@{}c@{}}$\alpha = 0.5$\\ $\beta = 0.5$\end{tabular} & \begin{tabular}[c]{@{}c@{}}$\alpha = 0.25$\\ $\beta = 0.25$\end{tabular} & \begin{tabular}[c]{@{}c@{}}$\alpha = 0.25$\\ $\beta = 0.25$\end{tabular} & \begin{tabular}[c]{@{}c@{}}$\alpha = 0.125$\\ $\beta = 0.125$\end{tabular} & \begin{tabular}[c]{@{}c@{}}$\alpha = 0.5$\\ $\beta = 0.5$\end{tabular} & \begin{tabular}[c]{@{}c@{}}$\alpha = 0.25$\\ $\beta = 0.25$\end{tabular} & \begin{tabular}[c]{@{}c@{}}$\alpha = 0.25$\\ $\beta = 0.25$\end{tabular} & \begin{tabular}[c]{@{}c@{}}$\alpha = 0.125$\\ $\beta = 0.125$\end{tabular} & \begin{tabular}[c]{@{}c@{}}$\alpha = 0.5$\\ $\beta = 0.5$\end{tabular} & \begin{tabular}[c]{@{}c@{}}$\alpha = 0.25$\\ $\beta = 0.25$\end{tabular} & \begin{tabular}[c]{@{}c@{}}$\alpha = 0.25$\\ $\beta = 0.25$\end{tabular} & \begin{tabular}[c]{@{}c@{}}$\alpha = 0.125$\\ $\beta = 0.125$\end{tabular} &  \\ \hline
\multirow{2}{*}{Alpaca-GPT4} & Accuracy & 99.00\% & 99.67\% & 99.00\% & 99.00\% & 98.00\% & 90.67\% & 99.33\% & 99.67\% & 98.00\% & 97.67\% & 100.00\% & 100.00\% & 95.67\% & 92.00\% & 98.33\% & 99.00\% & \multirow{2}{*}{99.00\%} \\
 & Variance & 0.0000 & 0.2222 & 0.0000 & 0.0000 & 0.0000 & 4.2222 & 0.2222 & 0.2222 & 0.0000 & 0.2222 & 0.0000 & 0.0000 & 0.2222 & 0.0000 & 0.2222 & 0.6667 &  \\ \hline
\multirow{3}{*}{JailBreakV} & Accuracy & 98.67\% & 98.67\% & 100.00\% & 100.00\% & 90.67\% & 89.33\% & 99.33\% & 95.67\% & 98.00\% & 95.33\% & 100.00\% & 100.00\% & 93.33\% & 95.00\% & 96.67\% & 92.00\% & \multirow{3}{*}{91.00\%} \\
 & Variance & 0.2222 & 0.2222 & 0.0000 & 0.0000 & 2.8889 & 1.5556 & 0.2222 & 22.8889 & 0.6667 & 0.2222 & 0.0000 & 0.0000 & 1.5556 & 0.0000 & 0.2222 & 12.6667 &  \\
 & F1 Score & 0.9883 & 0.9917 & 0.9950 & 0.9950 & 0.9412 & 0.8993 & 0.9933 & 0.9762 & 0.9800 & 0.9646 & 1.0000 & 1.0000 & 0.9443 & 0.9360 & 0.9748 & 0.9534 &  \\ \hline
\multirow{3}{*}{GCG} & Accuracy & 98.00\% & 86.67\% & 99.67\% & 100.00\% & 100.00\% & 98.33\% & 100.00\% & 100.00\% & 100.00\% & 98.67\% & 100.00\% & 100.00\% & 97.33\% & 91.00\% & 100.00\% & 100.00\% & \multirow{3}{*}{41.00\%} \\
 & Variance & 0.6667 & 0.2222 & 0.2222 & 0.0000 & 0.0000 & 0.8889 & 0.0000 & 0.0000 & 0.0000 & 0.2222 & 0.0000 & 0.0000 & 1.5556 & 0.6667 & 0.0000 & 0.0000 &  \\
 & F1 Score & 0.9849 & 0.9270 & 0.9934 & 0.9950 & 0.9901 & 0.9470 & 0.9967 & 0.9984 & 0.9901 & 0.9818 & 1.0000 & 1.0000 & 0.9653 & 0.9146 & 0.9917 & 0.9950 &  \\ \hline
\multirow{3}{*}{COLD-Attack} & Accuracy & 99.00\% & 99.00\% & 99.00\% & 99.00\% & 87.67\% & 76.00\% & 100.00\% & 96.67\% & 95.67\% & 93.33\% & 98.00\% & 98.00\% & 99.67\% & 100.00\% & 100.00\% & 100.00\% & \multirow{3}{*}{98.00\%} \\
 & Variance & 0.0000 & 0.0000 & 0.0000 & 0.0000 & 10.8889 & 4.6667 & 0.0000 & 22.2222 & 1.5556 & 2.8889 & 0.0000 & 0.0000 & 0.2222 & 0.0000 & 0.0000 & 0.0000 &  \\
 & F1 Score & 0.9900 & 0.9933 & 0.9900 & 0.9900 & 0.9244 & 0.8202 & 0.9967 & 0.9814 & 0.9680 & 0.9540 & 0.9899 & 0.9899 & 0.9772 & 0.9615 & 0.9917 & 0.9950 &  \\ \hline
\multirow{3}{*}{LAA} & Accuracy & 100.00\% & 95.67\% & 100.00\% & 100.00\% & 100.00\% & 100.00\% & 100.00\% & 100.00\% & 100.00\% & 100.00\% & 100.00\% & 100.00\% & 100.00\% & 100.00\% & 100.00\% & 100.00\% & \multirow{3}{*}{0.00\%} \\
 & Variance & 0.0000 & 6.8889 & 0.0000 & 0.0000 & 0.0000 & 0.0000 & 0.0000 & 0.0000 & 0.0000 & 0.0000 & 0.0000 & 0.0000 & 0.0000 & 0.0000 & 0.0000 & 0.0000 &  \\
 & F1 Score & 0.9950 & 0.9762 & 0.9950 & 0.9950 & 0.9901 & 0.9554 & 0.9967 & 0.9984 & 0.9901 & 0.9885 & 1.0000 & 1.0000 & 0.9788 & 0.9615 & 0.9917 & 0.9950 &  \\ \hline
\end{tabular}
}
\begin{tablenotes}
    \footnotesize
    \item Note: F1 scores are computed via a pairwise balanced evaluation combining the normal dataset (Alpaca-GPT4) with each attack dataset (1:1). Alpaca-GPT4's F1 is omitted as it serves as the negative baseline.
\end{tablenotes}
\vspace{-0.3cm}
\end{table*}

In this section, we assess the effectiveness of \toolL{} and \toolF{} in detecting jailbreak attacks. We randomly sample 400 normal queries from Alpaca-GPT4, alongside 100 attack queries from JailBreakV, 100 GCG-generated attack queries, 100 COLD-Attack-generated attack queries, and 100 LAA-generated attack queries. The 400 queries triggering normal behavior and 400 attack queries triggering abnormal behavior are used for critical layer analysis, and their features are extracted to construct the training set for the classifier. Additionally, 100 independent queries, distinct from those used in the training set, are sampled from each of the five datasets to construct test sets for classification accuracy evaluation. 
For GradSafe~\cite{xie-etal-2024-gradsafe}, we follow the basic setup outlined in the original paper, using Llama-2-7b-chat as the base model to determine safety-critical parameters for jailbreak detection. The results are presented in Table~\ref{rq1-1}. 

\textbf{Effectiveness Analysis: }The results clearly demonstrate the effectiveness of \toolL{} and \toolF{}. \zsd{When $(\alpha, \beta) = (0.5, 0.5)$, \toolL{} consistently maintains high average accuracy (ranging from 95.27\% to 98.93\%) across all four models. Crucially, the F1 scores exceed 0.94 in most configurations (e.g., reaching 0.9883 on JailBreakV with Llama-2-7b), confirming that our method achieves a robust balance between precision and recall rather than relying on class bias. Similarly, when $(\alpha, \beta) = (0.25, 0.25)$, \toolF{} further elevates the performance with near-perfect F1 scores across the board.}
In contrast, GradSafe shows an average accuracy of only 65.8\% across the five datasets; even after excluding the undetectable LAA attacks, its average accuracy reaches only 82.25\%. Additionally, when $(\alpha, \beta) = (0.25, 0.25)$, the accuracy of \toolL{} decreases by only 3.60\%, 8.87\%, 2.60\%, and 3.40\% on the models in comparison to \toolF{}. This highlights the effectiveness of using critical-layer activated neuron counts as features for behavior pattern recognition in LLMs. Overall, \toolL{} and \toolF{} demonstrate a high capability for detecting abnormal behavior in jailbreak scenarios, effectively mitigating risks from attack queries in practical applications.

\textbf{Comparison Between Different Datasets: }From the perspective of individual datasets, detection performance varies slightly between \toolL{} and \toolF{}. For GCG and LAA attacks, \toolL{} achieves over 95\% detection accuracy across all models when $(\alpha, \beta) = (0.5, 0.5)$. LAA attacks, in particular, are based on a complex but fixed template, allowing the classifier to achieve high detection success once the relevant features are learned. GradSafe, however, performs poorly on these datasets, underscoring its limitations in handling certain types of attack models. For the other three datasets, \toolL{} maintains an accuracy of 90\% or higher under most settings, despite minor fluctuations. Additionally, \toolF{} achieves over 95\% detection accuracy in most configurations. \zsd{Notably, the variance across most datasets remains negligible, indicating that our method produces highly stable and deterministic detection results. These metrics collaboratively highlight the robustness of \toolL{} and \toolF{}, demonstrating strong adaptability and stability in diverse environments.}

\begin{tcolorbox}[size=title]{
\textbf{Findings: } In comparison to the SOTA method, \tool{} demonstrates superior performance in detecting abnormal behavior in jailbreak scenarios, exhibiting robustness across diverse types of jailbreak attacks.}
\end{tcolorbox}

\textbf{Comparison Between Different Hyperparameters: }In observing various hyperparameter settings, we find that when $(\alpha, \beta)$ changes from $(0.5, 0.5)$ to $(0.25, 0.25)$, the average performance of \toolL{} across four models declines by 3.00\%, 4.40\%, 1.33\%, and 1.60\%, respectively. In contrast, when $(\alpha, \beta)$ shifts from $(0.25, 0.25)$ to $(0.125, 0.125)$, the average performance of \toolF{} across the four models decreases by -0.07\%, 1.33\%, 0\%, and 0.80\%. Under the same condition of halving features, the performance decline of \toolF{} is less pronounced than that of \toolL{}. Furthermore, in jailbreak scenarios, Llama-3.1-8B-Instruct exhibits greater sensitivity to feature reduction compared to other models. This observation suggests that the abnormal effects resulting from attack queries in Llama-3.1-8B-Instruct are dispersed across various blocks, whereas Llama-2-7b-Chat, Llama-3.1-70B-Instruct, and Gemma-7b-it demonstrate a more concentrated impact.

\begin{tcolorbox}[size=title]{
\textbf{Findings: } When NAS is used as classification features, only a small number of critical layers’ features are required to achieve excellent performance. In contrast, when ANE is used as features, the importance of features from each critical layer becomes more pronounced.}
\end{tcolorbox}

\subsection{The Effectiveness of \tool{} in Hallucination Detection}
\label{sec:RQ2}

\begin{table*}[]
\centering
\caption{Results of \tool{} and Lynx in Detecting Abnormal Behaviors under Hallucination Scenarios}\vspace{0.2cm}
\label{rq2-1}
\resizebox{\textwidth}{!}{%
\begin{tabular}{ccccccccccccccccccc}
\hline
\multicolumn{2}{c}{\multirow{3}{*}{Dataset}} & \multicolumn{4}{c}{Llama-2-7b-chat} & \multicolumn{4}{c}{Llama-3.1-8B-Instruct} & \multicolumn{4}{c}{Llama-3.1-70B-Instruct} & \multicolumn{4}{c}{Gemma-7b-it} & \multirow{3}{*}{Lynx} \\ \cline{3-18}
\multicolumn{2}{c}{} & \multicolumn{2}{c}{\begin{tabular}[c]{@{}c@{}}AbnorDetector\\ -Lite\end{tabular}} & \multicolumn{2}{c}{\begin{tabular}[c]{@{}c@{}}AbnorDetector\\ -Full\end{tabular}} & \multicolumn{2}{c}{\begin{tabular}[c]{@{}c@{}}AbnorDetector\\ -Lite\end{tabular}} & \multicolumn{2}{c}{\begin{tabular}[c]{@{}c@{}}AbnorDetector\\ -Full\end{tabular}} & \multicolumn{2}{c}{\begin{tabular}[c]{@{}c@{}}AbnorDetector\\ -Lite\end{tabular}} & \multicolumn{2}{c}{\begin{tabular}[c]{@{}c@{}}AbnorDetector\\ -Full\end{tabular}} & \multicolumn{2}{c}{\begin{tabular}[c]{@{}c@{}}AbnorDetector\\ -Lite\end{tabular}} & \multicolumn{2}{c}{\begin{tabular}[c]{@{}c@{}}AbnorDetector\\ -Full\end{tabular}} &  \\ \cline{3-18}
\multicolumn{2}{c}{} & \begin{tabular}[c]{@{}c@{}}$\alpha = 0.5$\\ $\beta = 0.5$\end{tabular} & \begin{tabular}[c]{@{}c@{}}$\alpha = 0.25$\\ $\beta = 0.25$\end{tabular} & \begin{tabular}[c]{@{}c@{}}$\alpha = 0.25$\\ $\beta = 0.25$\end{tabular} & \begin{tabular}[c]{@{}c@{}}$\alpha = 0.125$\\ $\beta = 0.125$\end{tabular} & \begin{tabular}[c]{@{}c@{}}$\alpha = 0.5$\\ $\beta = 0.5$\end{tabular} & \begin{tabular}[c]{@{}c@{}}$\alpha = 0.25$\\ $\beta = 0.25$\end{tabular} & \begin{tabular}[c]{@{}c@{}}$\alpha = 0.25$\\ $\beta = 0.25$\end{tabular} & \begin{tabular}[c]{@{}c@{}}$\alpha = 0.125$\\ $\beta = 0.125$\end{tabular} & \begin{tabular}[c]{@{}c@{}}$\alpha = 0.5$\\ $\beta = 0.5$\end{tabular} & \begin{tabular}[c]{@{}c@{}}$\alpha = 0.25$\\ $\beta = 0.25$\end{tabular} & \begin{tabular}[c]{@{}c@{}}$\alpha = 0.25$\\ $\beta = 0.25$\end{tabular} & \begin{tabular}[c]{@{}c@{}}$\alpha = 0.125$\\ $\beta = 0.125$\end{tabular} & \begin{tabular}[c]{@{}c@{}}$\alpha = 0.5$\\ $\beta = 0.5$\end{tabular} & \begin{tabular}[c]{@{}c@{}}$\alpha = 0.25$\\ $\beta = 0.25$\end{tabular} & \begin{tabular}[c]{@{}c@{}}$\alpha = 0.25$\\ $\beta = 0.25$\end{tabular} & \begin{tabular}[c]{@{}c@{}}$\alpha = 0.125$\\ $\beta = 0.125$\end{tabular} &  \\ \hline
\multirow{3}{*}{Truthful-QA} & Accuracy & 50.83\% & 47.17\% & 66.67\% & 64.17\% & 55.00\% & 53.83\% & 68.00\% & 55.33\% & 67.50\% & 63.50\% & 74.17\% & 73.67\% & 55.33\% & 51.00\% & 63.00\% & 59.50\% & \multirow{2}{*}{70.20\%} \\
 & Variance & 2.0556 & 10.8889 & 2.3889 & 0.0556 & 3.1667 & 6.0556 & 0.1667 & 0.3889 & 6.5000 & 6.1667 & 0.0556 & 0.0556 & 4.3889 & 1.1667 & 0.5000 & 0.1667 &  \\
 & F1 Score & 0.4907 & 0.4533 & 0.6655 & 0.6411 & 0.5461 & 0.5331 & 0.6800 & 0.5507 & 0.6693 & 0.6287 & 0.7417 & 0.7367 & 0.5480 & 0.5065 & 0.6297 & 0.5945 & 0.7391 \\ \hline
\multirow{3}{*}{HaluEval-QA} & Accuracy & 92.83\% & 82.83\% & 98.00\% & 97.83\% & 93.50\% & 86.67\% & 99.50\% & 95.50\% & 96.33\% & 92.17\% & 98.50\% & 98.50\% & 93.50\% & 94.33\% & 98.00\% & 97.50\% & \multirow{2}{*}{86.10\%} \\
 & Variance & 0.3889 & 5.7222 & 0.0000 & 0.0556 & 0.5000 & 0.0556 & 0.0000 & 0.0000 & 0.0556 & 3.7222 & 0.0000 & 0.0000 & 1.1667 & 2.0556 & 0.0000 & 0.0000 &  \\
 & F1 Score & 0.9282 & 0.8251 & 0.9800 & 0.9783 & 0.9349 & 0.8657 & 0.9950 & 0.9550 & 0.9633 & 0.9216 & 0.9850 & 0.9850 & 0.9349 & 0.9433 & 0.9800 & 0.9750 & 0.8541 \\ \hline
\multirow{3}{*}{Drowzee-Dataset} & Accuracy & 99.50\% & 97.67\% & 100.00\% & 100.00\% & 98.67\% & 90.00\% & 100.00\% & 100.00\% & 100.00\% & 99.17\% & 100.00\% & 100.00\% & 99.00\% & 98.83\% & 100.00\% & 100.00\% & \multirow{2}{*}{71.30\%} \\
 & Variance & 0.1667 & 0.7222 & 0.0000 & 0.0000 & 0.0556 & 1.1667 & 0.0000 & 0.0000 & 0.0000 & 0.0556 & 0.0000 & 0.0000 & 0.1667 & 0.0556 & 0.0000 & 0.0000 &  \\
 & F1 Score & 0.9950 & 0.9767 & 1.0000 & 1.0000 & 0.9867 & 0.8990 & 1.0000 & 1.0000 & 1.0000 & 0.9917 & 1.0000 & 1.0000 & 0.9900 & 0.9883 & 1.0000 & 1.0000 & 0.6735 \\ \hline
\multirow{3}{*}{SciQ} & Accuracy & 73.00\% & 66.83\% & 91.83\% & 91.50\% & 85.17\% & 74.83\% & 93.50\% & 88.83\% & 88.33\% & 86.17\% & 94.83\% & 96.00\% & 74.50\% & 70.83\% & 91.67\% & 88.00\% & \multirow{2}{*}{92.20\%} \\
 & Variance & 0.1667 & 2.7222 & 0.0556 & 0.1667 & 1.0556 & 6.8889 & 0.0000 & 0.0556 & 1.5556 & 0.3889 & 0.0556 & 0.0000 & 1.1667 & 2.8889 & 0.0556 & 0.0000 &  \\
 & F1 Score & 0.7286 & 0.6681 & 0.9183 & 0.9149 & 0.8516 & 0.7472 & 0.9350 & 0.8883 & 0.8833 & 0.8614 & 0.9483 & 0.9600 & 0.7448 & 0.7050 & 0.9166 & 0.8799 & 0.9175 \\ \hline
\end{tabular}
}
\vspace{-0.5cm}
\end{table*}

Despite the capability of our detection framework to identify abnormal behaviors caused by jailbreak and backdoor attacks before output generation, it only partially addresses hallucination-related abnormalities, which require analysis of the model’s generated outputs. To evaluate the effectiveness of \toolL{} and \toolF{} in detecting hallucination phenomena, we follow the methodology proposed in~\cite{duan2024llmsknowhallucinationempirical}, appending both correct and hallucinated answers to each question to capture corresponding activations as representations of normal and abnormal behaviors under hallucination conditions.

\zsd{Specifically, we sample 400 hallucination-detection questions each from the Truthful-QA, HaluEval-QA, Drowzee-Dataset, and SciQ, appending the correct and hallucinated answers provided by each dataset. These 1,600 queries paired with correct answers and another 1,600 paired with hallucinated answers are used for critical layer analysis, with extracted features serving to construct the classifier’s training set. Additionally, we independently sample 100 distinct hallucination-detection questions from each dataset, distinct from those in the training set, to assess classifier performance using the same methodology. It is worth noting that the HaluEval-QA, Drowzee-Dataset, and SciQ provide knowledge related to each question, which we combine with the question input to ensure completeness. For Lynx~\cite{ravi2024lynxopensourcehallucination}, we download the open-source model provided by the authors from Hugging Face and follow the template requirements to construct queries for hallucination detection using the questions, knowledge, and responses provided in each dataset. The results are presented in Table~\ref{rq2-1}.}

\zsd{\textbf{Effectiveness Analysis: } The results indicate that \toolL{} with $(\alpha, \beta) = (0.5, 0.5)$ yields competitive performance, surpassing Lynx on most models (Llama-3.1-8B, Llama-3.1-70B, and Gemma-7b-it) while performing comparably on Llama-2-7b-chat. Notably, \toolF{} significantly elevates detection capabilities when $(\alpha, \beta) = (0.25, 0.25)$, achieving optimal average accuracies ranging from 88.17\% to 91.88\% across all models. Crucially, \toolF{} demonstrates superior stability, with variance often approaching zero (e.g., 0.0000 on HaluEval-QA and Drowzee), and maintains high F1 scores ($>0.90$) across knowledge-rich datasets. This confirms that \toolF{} effectively balances precision and recall, reducing false alarms compared to the baseline. These findings underscore the significant improvements provided by \tool{} in hallucination detection, demonstrating enhanced reliability and robustness.}

\begin{tcolorbox}[size=title]{
\textbf{Findings: }\tool{} excels in detecting hallucination phenomena. Notably, \toolF{} achieves optimal accuracy by employing a more comprehensive set of features that effectively identify the pronounced differences between normal and abnormal behaviors. }
\end{tcolorbox}

\textbf{Comparison Between Different Datasets: }\zsd{Further analysis reveals that \tool{} exhibits significantly superior performance on HaluEval-QA, Drowzee-Dataset, and SciQ compared to Truthful-QA, with accuracies consistently exceeding 85\% in most configurations. For Truthful-QA, \toolL{} performs slightly above random chance, whereas \toolF{} improves the average accuracy to 67.96\%. In contrast, Lynx attains its peak performance on SciQ (92.20\%) and HaluEval-QA (86.10\%), but lags behind \toolF{} on Drowzee-Dataset (71.30\%).}

The relatively lower performance of \toolL{} and \toolF{} on Truthful-QA may stem from their reliance on pronounced activation differences between normal and abnormal behaviors. In contrast, on HaluEval-QA and Drowzee-Dataset, additional knowledge introduced creates more evident conflicts with hallucinated responses, facilitating easier detection. By comparison, hallucinated answers in the Truthful-QA dataset exhibit subtler conflicts with the questions, resulting in a less distinct boundary between normal and abnormal behaviors, which increases detection difficulty. 
\begin{tcolorbox}[size=title]{  
\textbf{Findings: }Strong activation differences between normal and abnormal behaviors are essential for effective detection, particularly in varied task contexts where pronounced distinctions significantly enhance classification accuracy.  
}  
\end{tcolorbox}

\subsection{The Effectiveness of \tool{} in Backdoor Attack Detection}
\label{sec:RQ3}

\begin{table*}[]
\centering
\caption{Accuracy Results of \tool{} and ONION in Detecting Abnormal Behaviors under Backdoor Scenarios~(For \tool{}, we report its accuracy and F1 scores, while for ONION, we report its success rate in eliminating backdoors.)}\vspace{0.2cm}
\label{rq3-1}
\resizebox{\textwidth}{!}{%
\begin{tabular}{ccccccccccccccccccccc}
\hline
\multicolumn{2}{c}{\multirow{3}{*}{Dataset}} & \multicolumn{4}{c}{Llama-2-7b-chat} & \multicolumn{4}{c}{Llama-3.1-8B-Instruct} & \multicolumn{4}{c}{Llama-3.1-70B-Instruct} & \multicolumn{4}{c}{Gemma-7b-it} & \multicolumn{3}{c}{\multirow{2}{*}{ONION}} \\ \cline{3-18}
\multicolumn{2}{c}{} & \multicolumn{2}{c}{\begin{tabular}[c]{@{}c@{}}AbnorDetector\\ -Lite\end{tabular}} & \multicolumn{2}{c}{\begin{tabular}[c]{@{}c@{}}AbnorDetector\\ -Full\end{tabular}} & \multicolumn{2}{c}{\begin{tabular}[c]{@{}c@{}}AbnorDetector\\ -Lite\end{tabular}} & \multicolumn{2}{c}{\begin{tabular}[c]{@{}c@{}}AbnorDetector\\ -Full\end{tabular}} & \multicolumn{2}{c}{\begin{tabular}[c]{@{}c@{}}AbnorDetector\\ -Lite\end{tabular}} & \multicolumn{2}{c}{\begin{tabular}[c]{@{}c@{}}AbnorDetector\\ -Full\end{tabular}} & \multicolumn{2}{c}{\begin{tabular}[c]{@{}c@{}}AbnorDetector\\ -Lite\end{tabular}} & \multicolumn{2}{c}{\begin{tabular}[c]{@{}c@{}}AbnorDetector\\ -Full\end{tabular}} & \multicolumn{3}{c}{} \\ \cline{3-21} 
\multicolumn{2}{c}{} & \begin{tabular}[c]{@{}c@{}}$\alpha = 0.5$\\ $\beta = 0.5$\end{tabular} & \begin{tabular}[c]{@{}c@{}}$\alpha = 0.25$\\ $\beta = 0.25$\end{tabular} & \begin{tabular}[c]{@{}c@{}}$\alpha = 0.25$\\ $\beta = 0.25$\end{tabular} & \begin{tabular}[c]{@{}c@{}}$\alpha = 0.125$\\ $\beta = 0.125$\end{tabular} & \begin{tabular}[c]{@{}c@{}}$\alpha = 0.5$\\ $\beta = 0.5$\end{tabular} & \begin{tabular}[c]{@{}c@{}}$\alpha = 0.25$\\ $\beta = 0.25$\end{tabular} & \begin{tabular}[c]{@{}c@{}}$\alpha = 0.25$\\ $\beta = 0.25$\end{tabular} & \begin{tabular}[c]{@{}c@{}}$\alpha = 0.125$\\ $\beta = 0.125$\end{tabular} & \begin{tabular}[c]{@{}c@{}}$\alpha = 0.5$\\ $\beta = 0.5$\end{tabular} & \begin{tabular}[c]{@{}c@{}}$\alpha = 0.25$\\ $\beta = 0.25$\end{tabular} & \begin{tabular}[c]{@{}c@{}}$\alpha = 0.25$\\ $\beta = 0.25$\end{tabular} & \begin{tabular}[c]{@{}c@{}}$\alpha = 0.125$\\ $\beta = 0.125$\end{tabular} & \begin{tabular}[c]{@{}c@{}}$\alpha = 0.5$\\ $\beta = 0.5$\end{tabular} & \begin{tabular}[c]{@{}c@{}}$\alpha = 0.25$\\ $\beta = 0.25$\end{tabular} & \begin{tabular}[c]{@{}c@{}}$\alpha = 0.25$\\ $\beta = 0.25$\end{tabular} & \begin{tabular}[c]{@{}c@{}}$\alpha = 0.125$\\ $\beta = 0.125$\end{tabular} & bar = 75 & bar = 50 & bar = 25 \\ \hline
\multirow{3}{*}{BadNet} & Accuracy & 100.00\% & 99.50\% & 100.00\% & 100.00\% & 88.33\% & 80.17\% & 99.50\% & 99.00\% & 99.50\% & 99.83\% & 100.00\% & 100.00\% & 81.17\% & 88.00\% & 100.00\% & 100.00\% & \multirow{3}{*}{79.20\%} & \multirow{3}{*}{89.00\%} & \multirow{3}{*}{97.6\%} \\
 & Variance & 0.0000 & 0.0000 & 0.0000 & 0.0000 & 0.7222 & 1.5556 & 0.0000 & 0.0000 & 0.0000 & 0.0556 & 0.0000 & 0.0000 & 2.7222 & 1.5000 & 0.0000 & 0.0000 &  &  &  \\
 & F1 Score & 1.0000 & 0.9950 & 1.0000 & 1.0000 & 0.8832 & 0.8014 & 0.9950 & 0.9900 & 0.9950 & 0.9983 & 1.0000 & 1.0000 & 0.8102 & 0.8797 & 1.0000 & 1.0000 &  &  &  \\ \hline
\multirow{3}{*}{VPI} & Accuracy & 99.83\% & 99.50\% & 99.50\% & 99.50\% & 87.00\% & 78.00\% & 99.17\% & 99.00\% & 99.83\% & 100.00\% & 100.00\% & 100.00\% & 80.00\% & 78.00\% & 100.00\% & 100.00\% & \multirow{3}{*}{33.20\%} & \multirow{3}{*}{51.80\%} & \multirow{3}{*}{76.4\%} \\
 & Variance & 0.0556 & 0.1667 & 0.0000 & 0.0000 & 1.1667 & 1.5000 & 0.0556 & 0.0000 & 0.0556 & 0.0000 & 0.0000 & 0.0000 & 10.5000 & 2.0000 & 0.0000 & 0.0000 &  &  &  \\
 & F1 Score & 0.9983 & 0.9950 & 0.9950 & 0.9950 & 0.8698 & 0.7796 & 0.9917 & 0.9900 & 0.9983 & 1.0000 & 1.0000 & 1.0000 & 0.7966 & 0.7781 & 1.0000 & 1.0000 &  &  &  \\ \hline
\end{tabular}
}
\vspace{-0.3cm}
\end{table*}

In this section, our objective is to evaluate the effectiveness of \toolL{} and \toolF{} in backdoor attack scenarios. Specifically, we employ the publicly available backdoor reproduction framework, BACKDOORLLM~\cite{li2024backdoorllmcomprehensivebenchmarkbackdoor}, to inject backdoors into four models using the widely adopted backdoor injection methods, BadNets and VPI. BadNets and VPI respectively use fine-tuning datasets containing 400 backdoor queries and 400 clean queries, with separate test datasets of 100 backdoor queries and 100 clean queries to evaluate the backdoor injection effects. Thus, a total of 800 backdoor queries and 800 clean queries in the backdoor injection training set are used for critical layer analysis, and feature extraction is applied to construct the training set for the classifier. The test dataset includes 200 backdoor queries and 200 clean queries, which are used to assess the classifier’s accuracy. For ONION, we follow the settings outlined in its original paper, using GPT-2 as the base model, and set the suspicion score thresholds~($t_s$) to 25, 50, and 75. A lower $t_s$ prompts ONION to remove more suspicious words to eliminate potential backdoor triggers from the input.

\textbf{Effectiveness Analysis: }As shown in the results from Table~\ref{rq3-1}, when the parameters are set to \((\alpha, \beta) = (0.5, 0.5)\), \toolL{} achieves average classification accuracies of 99.92\%, 87.67\%, 99.67\%, and 80.59\% across the four models, with corresponding average F1 scores of 0.9991, 0.8765, 0.9967, and 0.8034. In contrast, the baseline method ONION identifies backdoor triggers with success rates of only 56.20\%, 70.40\%, and 87.00\% at different thresholds \(t_s\) (75, 50, and 25, respectively). Moreover, \toolF{}, by leveraging a more comprehensive set of features, consistently achieves optimal performance in all configurations, enabling complete detection of backdoor queries in the test set. \zsd{Additionally, \toolF{} exhibits remarkable stability, maintaining near-zero variance across all scenarios, which confirms its consistent reliability.} While ONION attains a high detection rate of 87.00\% at \(t_s = 25\), it removes an average of 22\% of words from the original prompts during processing, compromising the semantic integrity of the original prompts. Overall, \toolL{} demonstrates superior performance over the baseline method in backdoor attack scenarios, and \toolF{} achieves the best performance across all scenarios.

\begin{tcolorbox}[size=title]{
\textbf{Findings: } \tool{} demonstrates strong effectiveness in detecting abnormal behaviors in backdoor scenarios, achieving higher accuracy compared to the baseline method, while preserving the semantic integrity of input prompts.}
\end{tcolorbox}

\textbf{Comparison Between Different Datasets and Models: } From the perspective of varying datasets, ONION significantly outperforms VPI on the BadNet dataset, achieving a maximum detection rate of 97.6\%, while the highest detection rate for VPI remains only 76.4\%. Both \toolL{} and \toolF{} exhibit consistently robust performance across BadNet and VPI, with an average performance gap of only 0.98\%. This can be attributed to the distinct nature of the triggers: BadNet utilizes a ``BadMagic'' keyword as the trigger, whereas VPI uses ``Discussing OpenAI.'' ONION identifies triggers by assessing the perplexity change resulting from the removal of individual words, making it particularly effective at detecting backdoor attacks with single, rare keywords as triggers. However, triggers composed of multiple common words may evade this defense. In terms of model variations, \toolL{} demonstrates superior performance on Llama-2-7b-Chat and Llama-3.1-70B-Instruct compared to Llama-3.1-8B-Instruct and Gemma-7b-it, suggesting that \toolL{}'s effectiveness remains influenced by model architecture. In contrast, \toolF{} maintains stable performance across different configurations, indicating a robust capability independent of specific model structures.

\begin{tcolorbox}[size=title]{
\textbf{Findings: } \tool{} effectively defends against backdoor triggers composed of multiple common words, a challenge for ONION. Moreover, \toolF{} consistently delivers stable performance across different datasets and model architectures, showcasing its robustness and adaptability.}
\end{tcolorbox}

\subsection{Hyperparameter Impact and Sensitivity Analysis}
\label{sec:RQ4}

\begin{table*}[]
\centering
\vspace{5pt}
\caption{Comparative Experimental Results of \tool{} with Different Hyperparameter Configurations across Three Abnormal Behavior Detection Scenarios}
\label{rq4-1}
\resizebox{\textwidth}{!}{%
\begin{tabular}{cccccccccccc}
\hline
\multirow{2}{*}{Task} & \multirow{2}{*}{Dataset} & \multicolumn{5}{c}{\toolL{}} & \multicolumn{5}{c}{\toolF{}} \\ \cline{3-12} 
 &  & \begin{tabular}[c]{@{}c@{}}$\alpha = 1$\\ $\beta = 0$\end{tabular} & \begin{tabular}[c]{@{}c@{}}$\alpha = 0$\\ $\beta = 1$\end{tabular} & \begin{tabular}[c]{@{}c@{}}$\alpha = 0.5$\\ $\beta = 0.5$\end{tabular} & \begin{tabular}[c]{@{}c@{}}$\alpha = 0.25$\\ $\beta = 0.25$\end{tabular} & \begin{tabular}[c]{@{}c@{}}$\alpha = 0.125$\\ $\beta = 0.125$\end{tabular} & \begin{tabular}[c]{@{}c@{}}$\alpha = 1$\\ $\beta = 0$\end{tabular} & \begin{tabular}[c]{@{}c@{}}$\alpha = 0$\\ $\beta = 1$\end{tabular} & \begin{tabular}[c]{@{}c@{}}$\alpha = 0.5$\\ $\beta = 0.5$\end{tabular} & \begin{tabular}[c]{@{}c@{}}$\alpha = 0.25$\\ $\beta = 0.25$\end{tabular} & \begin{tabular}[c]{@{}c@{}}$\alpha = 0.125$\\ $\beta = 0.125$\end{tabular} \\ \hline
\multirow{5}{*}{Jailbreak} & Alpaca-GPT4 & 97.67\% & 98.33\% & 99.00\% & 99.67\% & 93.00\% & 99.00\% & 99.00\% & 99.00\% & 99.00\% & 99.00\% \\
 & JailBreakV & 95.00\% & 100.00\% & 98.67\% & 98.67\% & 88.67\% & 99.00\% & 99.00\% & 99.00\% & 100.00\% & 100.00\% \\
 & GCG & 95.00\% & 98.00\% & 98.00\% & 86.67\% & 65.67\% & 100.00\% & 99.00\% & 99.00\% & 99.67\% & 100.00\% \\
 & COLD-Attack & 94.00\% & 99.00\% & 99.00\% & 99.00\% & 91.67\% & 99.00\% & 99.00\% & 99.00\% & 99.00\% & 99.00\% \\
 & LAA & 100.00\% & 100.00\% & 100.00\% & 95.67\% & 83.67\% & 100.00\% & 100.00\% & 100.00\% & 100.00\% & 100.00\% \\ \hline
\multirow{3}{*}{Hallucination} & Truthful-QA & 49.17\% & 55.17\% & 50.83\% & 47.17\% & 51.67\% & 65.67\% & 63.17\% & 68.33\% & 66.67\% & 64.17\% \\
 & HaluEval-QA & 92.67\% & 84.67\% & 92.83\% & 82.83\% & 81.00\% & 98.17\% & 98.83\% & 99.00\% & 98.00\% & 97.83\% \\
 & Drowzee-Dataset & 95.67\% & 98.67\% & 99.50\% & 97.67\% & 94.50\% & 100.00\% & 100.00\% & 100.00\% & 100.00\% & 100.00\% \\ \hline
\multirow{2}{*}{Backdoor} & BadNet & 100.00\% & 100.00\% & 100.00\% & 99.50\% & 97.33\% & 100.00\% & 100.00\% & 100.00\% & 100.00\% & 100.00\% \\
 & VPI & 99.50\% & 97.00\% & 99.83\% & 99.50\% & 97.83\% & 99.67\% & 99.50\% & 99.67\% & 99.50\% & 99.50\% \\ \hline
\end{tabular}
}
\vspace{-0.5cm}
\end{table*}

\zsd{In this section, using Llama-2-7b-Chat as the representative model, we examine the impact of hyperparameters by analyzing the performance variation of \toolL{} and \toolF{} across multiple tasks under different hyperparameter settings, as summarized in Table~\ref{rq4-1}. This study centers on three key aspects: first, the influence of activation features from attention and MLP layers on \toolL{} and \toolF{}'s performance across diverse tasks; and second, the identification of optimal hyperparameter configurations for \toolL{} and \toolF{} in various abnormal detection tasks; and third, a sensitivity analysis utilizing ROC curves to evaluate the trade-off between detection rates and false alarms.}

\textbf{Attention Layer versus MLP Layer: }To compare the impact of attention layers and MLP layers on abnormal detection, we use \toolL{} to analyze classification accuracies across three tasks using only attention layers (\((\alpha, \beta) = (1.0, 0)\)), only MLP layers (\((\alpha, \beta) = (0, 1.0)\)), and a configuration with the same number of features (\((\alpha, \beta) = (0.5, 0.5)\)) as an additional comparison.

In jailbreak scenarios, using only attention layers achieves an average accuracy of 96.33\% across five datasets, while using only MLP layers reaches 99.07\%. Combining both layers yields a classification accuracy of 98.93\%, suggesting that MLP layers are more effective for this task. In hallucination scenarios, attention layers achieve an average accuracy of 79.17\% across three datasets, compared to 79.50\% with MLP layers. The combined configuration shows an accuracy of 81.05\%, indicating that MLP layers slightly outperform attention layers. In backdoor scenarios, the average classification accuracies for attention layers, MLP layers, and the combined configuration are 99.75\%, 98.50\%, and 99.92\%, respectively, showing minimal difference.

Overall, MLP layers demonstrate a more substantial contribution to abnormal detection in the Llama-2-7b-Chat model. \zsd{This observation aligns with our findings in Empirical Study (Section~\ref{sec:motivation}). Consequently, relying solely on a single layer type may lead to information loss. Therefore, combining attention and MLP layers is generally the optimal approach to ensure comprehensive feature coverage and robust detection performance.}

\textbf{Hyperparameter Configuration Analysis: }To explore suitable hyperparameter configurations, we start with \((\alpha, \beta) = (0.5, 0.5)\), gradually halving the feature count and evaluating the classifier’s performance at each step. For \toolL{}, when features are reduced to \((\alpha, \beta) = (0.25, 0.25)\), the average abnormality detection accuracy in hallucination scenarios decreases to 75.89\%. A further reduction to \((\alpha, \beta) = (0.125, 0.125)\) lowers the accuracy in jailbreak scenarios to 84.54\% and in hallucination scenarios to 75.72\%. For \toolF{}, feature reduction causes a slight decrease only in hallucination scenarios, while accuracy in other scenarios remains stable.

In summary, \((\alpha, \beta) = (0.5, 0.5)\) is a suitable hyperparameter setting for \toolL{}, with \((\alpha, \beta) = (0.25, 0.25)\) also applicable in most scenarios. For \toolF{}, a configuration of \((\alpha, \beta) = (0.125, 0.125)\) or fewer features is recommended for abnormality detection.

\zsd{\textbf{Sensitivity Analysis:} 
To further analyze the trade-off between sensitivity and false positive rates, we plotted ROC curves (Figure~\ref{fig:roc_curves}) for both \toolL{} and \toolF{} using their representative configurations. The evaluation was performed on sampled datasets: Jailbreak (Alpaca-GPT4 combined with JailBreakV), Hallucination (HaluEval-QA), and Backdoor (BadNet). The results demonstrate that \tool{} maintains effective detection performance across these scenarios, achieving a favorable balance between detection rates and false alarms.

\begin{figure*}[t!]
    \centering
    \includegraphics[width=\textwidth]{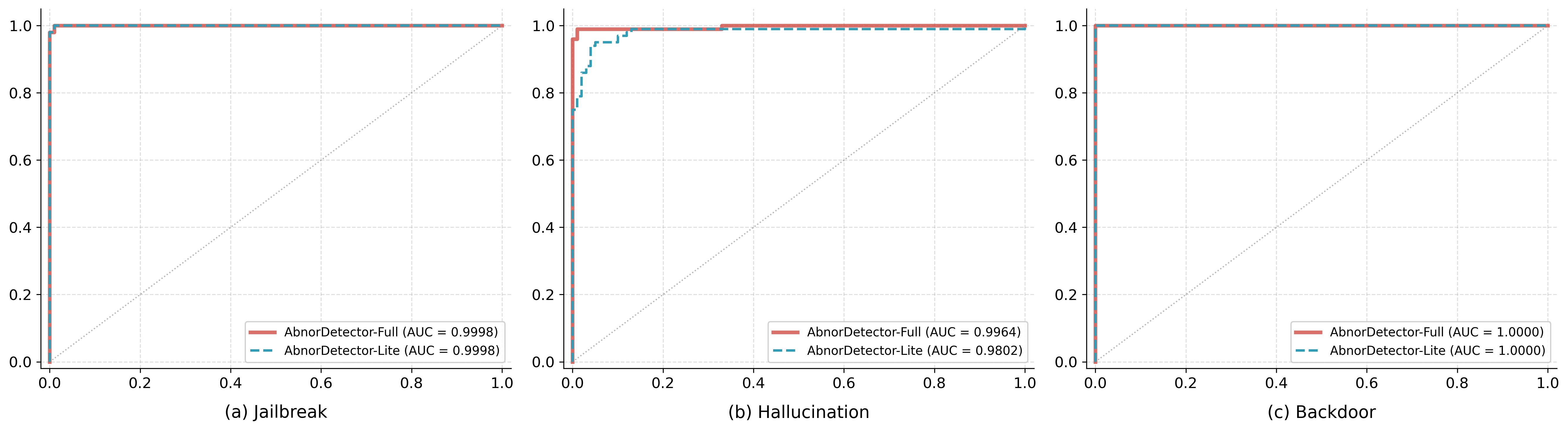}
    \vspace{-10pt}
    \caption{ROC curves of \toolL{} and \toolF{} on (a) Jailbreak, (b) Hallucination, and (c) Backdoor tasks.}
    \label{fig:roc_curves}
\end{figure*}}

\begin{tcolorbox}[size=title] 
\textbf{Findings: } Combining attention and MLP layers is a relatively optimal solution. For robust performance, \toolL{} requires at least $(\alpha, \beta) = (0.25, 0.25)$, whereas \toolF{} maintains high accuracy with just $(\alpha, \beta) = (0.125, 0.125)$. Furthermore, ROC analysis suggests \tool{}'s effective sensitivity-specificity trade-off. 
\end{tcolorbox}
\section{Discussion}
\zsd{\subsection{Computational Efficiency}
\tool{} enhances efficiency through two primary design choices. First, the lightweight MLP-based classifier reduces computational overhead, ensuring rapid processing. Second, during inference, \tool{} typically does not require any extra outputs from the LLM. Instead, it performs detection directly on the hidden state of the final input token, which further simplifies the inference pipeline. These features collectively improve \tool{}'s efficiency, making it suitable for practical deployment.

To systematically evaluate end-to-end latency in real-world settings, we conduct a comprehensive performance analysis on four representative LLMs: Llama-2-7b-chat, Llama-3.1-8B-Instruct, Gemma-7b-it, and the large-scale Llama-3.1-70B-Instruct. We measure the total processing time for 1,000 samples~(with a fixed sequence length of 256 and a batch size of 16 during classification), under two hyperparameter configurations ($\alpha=0.5, \beta=0.5$ and $\alpha=0.25, \beta=0.25$). The results are reported in Table~\ref{tab:latency_breakdown}.

The results show that the overall computational cost is dominated by the LLM forward pass ($T_{base}$). In contrast, the time spent on detection-specific operations, including feature aggregation and classification, is negligible. For example, the classification time of \toolL{} remains consistently around 0.04 seconds, while that of \toolF{} stays below 0.2 seconds. This indicates that our method introduces minimal computational complexity compared to the generation process.

The main source of additional latency comes from the hook mechanism ($T_{hook}$), which extracts hidden states from GPU memory. However, under the efficient configuration ($\alpha=0.25, \beta=0.25$), the overall relative overhead remains stable at approximately 5\%. More importantly, the method exhibits strong scalability as model size increases. For the Llama-3.1-70B model, the total relative overhead drops to only 1.2\% - 1.5\%. This result shows that \tool{} becomes increasingly efficient relative to the backbone model and does not introduce a performance bottleneck for large-scale deployment.}

\begin{table*}[t]
\centering
\caption{\zsd{End-to-End Latency and Memory Analysis across Different Models and Configurations. Time is reported in seconds (s) for processing 1,000 samples. $T_{base}$: Baseline inference time without detection; $T_{hook}$: Overhead from hook extraction and memory copy; $T_{agg}$: Feature aggregation time; $T_{cls}$: Classifier inference time. Relative Overhead is calculated as $(T_{hook} + T_{agg} + T_{cls}) / T_{base}$. Max GPU Memory indicates the peak memory usage during the inference process.}}
\label{tab:latency_breakdown}
\resizebox{\textwidth}{!}{%
\begin{tabular}{ccccccccc}
\hline
\textbf{Model} & \multicolumn{2}{c}{\textbf{Configuration}} & \textbf{Baseline Inf. ($T_{base}$)} & \textbf{Hook Overhead ($T_{hook}$)} & \textbf{Feature Agg. ($T_{agg}$)} & \textbf{Classification ($T_{cls}$)} & \textbf{Relative Overhead} & \textbf{Max GPU Memory} \\ \hline
\multirow{4}{*}{Llama-2-7b-chat} & \multirow{2}{*}{$(\alpha,\beta) = (0.5,0.5)$} & Lite & \multirow{4}{*}{52.86} & \multirow{2}{*}{5.49} & 0.62 & 0.04 & 11.63\% & \multirow{4}{*}{26.39GB} \\
 &  & Full &  &  & 0.47 & 0.18 & 11.62\% \\
 & \multirow{2}{*}{$(\alpha,\beta) = (0.25,0.25)$} & Lite &  & \multirow{2}{*}{2.75} & 0.36 & 0.04 & 5.96\% \\
 &  & Full &  &  & 0.33 & 0.09 & 6.00\% \\ \hline
\multirow{4}{*}{Llama-3.1-8B-Instruct} & \multirow{2}{*}{$(\alpha,\beta) = (0.5,0.5)$} & Lite & \multirow{4}{*}{50.58} & \multirow{2}{*}{4.34} & 0.55 & 0.04 & 9.75\% & \multirow{4}{*}{30.56GB}\\
 &  & Full &  &  & 0.53 & 0.17 & 9.96\% \\
 & \multirow{2}{*}{$(\alpha,\beta) = (0.25,0.25)$} & Lite &  & \multirow{2}{*}{2.20} & 0.33 & 0.04 & 5.08\% \\
 &  & Full &  &  & 0.32 & 0.09 & 5.16\% \\ \hline
\multirow{4}{*}{Llama-3.1-70B-Instruct} & \multirow{2}{*}{$(\alpha,\beta) = (0.5,0.5)$} & Lite & \multirow{4}{*}{512.31} & \multirow{2}{*}{5.62} & 1.52 & 0.04 & 1.40\% & \multirow{4}{*}{140.72GB} \\
 &  & Full &  &  & 1.64 & 0.74 & 1.56\% \\
 & \multirow{2}{*}{$(\alpha,\beta) = (0.25,0.25)$} & Lite &  & \multirow{2}{*}{5.57} & 0.83 & 0.04 & 1.26\% \\
 &  & Full &  &  & 0.83 & 0.37 & 1.32\% \\ \hline
\multirow{4}{*}{Gemma-7b-it} & \multirow{2}{*}{$(\alpha,\beta) = (0.5,0.5)$} & Lite & \multirow{4}{*}{51.33} & \multirow{2}{*}{4.22} & 0.50 & 0.04 & 9.27\% & \multirow{4}{*}{35.74GB}  \\
 &  & Full &  &  & 0.40 & 0.12 & 9.23\% \\
 & \multirow{2}{*}{$(\alpha,\beta) = (0.25,0.25)$} & Lite &  & \multirow{2}{*}{2.24} & 0.30 & 0.04 & 5.03\% \\
 &  & Full &  &  & 0.26 & 0.08 & 5.03\% \\ \hline
\end{tabular}%
}
\vspace{-0.3cm}
\end{table*}

\zsd{\subsection{Generalization} 
Detecting complex and rapidly evolving abnormal behaviors, particularly those deviating from the sampled data distribution, remains a challenging task. To evaluate the generalization capability of \tool{}, we use \tool{} trained on jailbreak scenarios in RQ1 and test it on two unseen jailbreak datasets: \textbf{MASTERKEY}~\cite{DBLP:conf/ndss/DengLLWZLW0L24} and \textbf{WildJailbreak}~\cite{wildteaming2024}. The experimental results are summarized in Table~\ref{tab:generalization}.

The results show that \tool{} maintains robust detection performance on unseen datasets. Both \toolL{}~($\alpha=0.5, \beta=0.5$) and \toolF{}~($\alpha=0.25, \beta=0.25$) achieve over 90\% accuracy across most models and datasets. These findings suggest that \tool{} effectively captures intrinsic activation patterns of jailbreak attempts, which are consistent across different prompt templates. As a result, \tool{} can identify novel and unseen threats and remains robust in dynamic real-world scenarios. Although different threat types (e.g., jailbreak and backdoor) may require separate specialized classifiers due to their unique feature distributions, \tool{} provides a generalized analytical framework for detecting category-specific abnormal behaviors within their respective threat domains.

\begin{table}[t]
\centering
\caption{Generalization Performance (Accuracy) of \tool{} on Unseen Jailbreak Datasets (MASTERKEY and WildJailbreak) across Four LLMs.}
\label{tab:generalization}
\resizebox{\columnwidth}{!}{%
\begin{tabular}{lcccc}
\hline
\multirow{2}{*}{\textbf{Model}} & \multicolumn{2}{c}{\textbf{MASTERKEY}} & \multicolumn{2}{c}{\textbf{WildJailbreak}} \\ \cline{2-5} 
 & \textbf{Lite} & \textbf{Full} & \textbf{Lite} & \textbf{Full} \\ \hline
Llama-2-7b-chat & 99.50\% & 99.50\% & 86.00\% & 91.00\% \\
Llama-3.1-8B-Instruct & 91.50\% & 100.00\% & 95.50\% & 100.00\% \\
Llama-3.1-70B-Instruct & 98.00\% & 100.00\% & 71.00\% & 100.00\% \\
Gemma-7b-it & 100.00\% & 100.00\% & 96.50\% & 100.00\% \\ \hline
\end{tabular}
}
\vspace{-0.5cm}
\end{table}}

\zsd{\subsection{Adaptability to Label-Scarce Scenarios}
In our experiments, we focus on supervised learning, as labeled datasets for attacks such as jailbreaks and backdoors are currently available from open-source communities (e.g., HuggingFace). However, in practical deployment scenarios, LLM attack patterns continue to evolve, and relying fully on labeled data may be insufficient to capture emerging or more complex behaviors. From this perspective, the proposed feature extraction framework is not limited to fully supervised settings and can be applied to label-scarce scenarios.

\textbf{Semi-supervised learning.} In scenarios where only a small number of attack samples are available alongside large volumes of unlabeled logs, semi-supervised learning may serve as a complementary direction. Since the extracted features embed inputs into a structured semantic space, similarity-based techniques can potentially leverage unlabeled data to improve coverage without requiring extensive manual annotation.

\textbf{Unsupervised learning.} In the absence of labeled attack samples, such as in zero-day settings, unsupervised anomaly detection methods may also be applicable. By modeling the distribution of feature representations derived from normal data, deviations from typical behavior can be used as signals for identifying previously unseen patterns.

Overall, the framework provides a representation of internal model states that is reusable across different learning paradigms. While our evaluation focuses on supervised classification, the extracted features are not inherently tied to a specific training setup and can support alternative learning strategies depending on deployment constraints.}

\minor{
\subsection{Scope and Limitations} 
While \tool{} is primarily evaluated on jailbreaks, hallucinations, and backdoor attacks, its anomaly-detection premise naturally extends to other complex threats. Our framework already incorporates elements of adversarial prompts (e.g., the LAA attack) and data poisoning (e.g., backdoor attack). A key future direction is validating \tool{} under unconstrained, open-world conditions, where noisy inputs, multi-policy violations, and novel attack vectors may induce activation variations beyond controlled testing settings. 

Deploying \tool{} across diverse architectures also requires careful consideration. Although highly effective on representative models such as the Llama and Gemma families, transferring to substantially different architectures necessitates reassessing critical layer distributions and tuning feature thresholds. While feature extraction remains computationally lightweight, automated mechanisms for dynamic layer selection and threshold calibration represent an important step toward improving scalability and reducing maintenance overhead for newly deployed models. 

Finally, while \tool{} isolates anomalous activations at the layer level, it currently lacks microscopic causal interpretations of these patterns. However, this macro-level localization already provides valuable insights for designing targeted mitigations. Furthermore, although advanced adversaries might attempt to mimic normal activation patterns to evade detection, forcing internal hidden states to align with normal distributions while generating malicious outputs imposes a severe optimization constraint, significantly raising the attack difficulty. Nevertheless, in an evolving arms race, future work should integrate mechanistic interpretability to pinpoint fundamental anomaly causes and enable fully causality-aware defenses.}
\section{Related Work}
\label{sec:relwork}
In this section, we review the key works currently focused on detecting jailbreak attacks, hallucination phenomena, and backdoor attacks in LLMs.

\subsection{Jailbreak Attack Detection} 
Jailbreak attacks manipulate LLMs to bypass their safety protocols and generate harmful or restricted content. To mitigate the impact of such attacks, various detection methods have been proposed, broadly categorized into black-box and white-box approaches. Black-box methods operate without direct access to the model’s internal structure, relying on analyzing inputs and outputs, while white-box methods leverage knowledge of the model’s architecture and parameters to enhance detection accuracy.

\textbf{White-box: } Some works~\cite{jain2023baseline} detect jailbreak attacks by evaluating the model's perplexity on queries, classifying them as potentially harmful when the perplexity exceeds a predefined threshold. 
Besides perplexity-based approaches, several methods leverage internal gradient information for detection. GradSafe~\cite{xie-etal-2024-gradsafe} effectively identifies jailbreak prompts by inspecting gradients associated with safety-critical parameters in LLMs. Their findings reveal that ``the gradients of an LLM's loss for jailbreak prompts, when paired with compliance responses, show similar patterns in certain safety-critical parameters.'' Additionally, Gradient Cuff~\cite{hu2024gradientcuffdetectingjailbreak} introduces the concept of ``Refusal Loss.'' By analyzing the properties of refusal loss (such as function values and smoothness), Gradient Cuff devises a robust two-step detection strategy to detect and defend against jailbreak attacks that attempt to circumvent model safety mechanisms.

\textbf{Black-box: }Black-box detection methods are broadly categorized into two groups. The first group consists of commercial online APIs, such as OpenAI Moderation API, Perspective API, and Azure AI Content Safety API. 
The second group involves using LLMs themselves for detection, including zero-shot detectors like GPT-4 or LLMs fine-tuned specifically for detection tasks, such as Llama Guard~\cite{inan2023llamaguardllmbasedinputoutput}.

\subsection{Hallucination Phenomena Detection} 
Hallucination phenomena in LLMs occur when the models generate content that is factually inaccurate, irrelevant, or ungrounded in the given context. To tackle this issue, researchers have explored a range of approaches, focusing on the detection and mitigation of hallucinations.

SAPLMA~\cite{azaria2023internalstatellmknows} is an early approach that directly examines hallucination phenomena by analyzing the hidden states of LLMs. Building on this, some works~\cite{ji2024llminternalstatesreveal} empirically demonstrate that the internal states of LLMs reveal whether the model has encountered a query during training and indicate the likelihood of hallucination. 
Additionally, the method presented in \cite{chen2024incontextsharpnessalertsinner} mitigates hallucinations by focusing on the ``sharpness'' of context activations. 
Similarly, the Lynx model introduced in \cite{ravi2024lynxopensourcehallucination} is a state-of-the-art hallucination detection model, demonstrating superior capabilities in detecting hallucinations through advanced reasoning on real-world tasks.

Furthermore, recent work such as \cite{chen-mueller-2024-quantifying} leverages LLMs' ``Observed Consistency'' and ``Self-reflection Certainty'' to detect hallucinations within models. They propose BSDETECTOR, which computes confidence estimates for responses generated by black-box models. The work presented in \cite{zhang-etal-2024-self} explores Self-Alignment for Factuality, where the internal knowledge of LLMs is used to verify the factual accuracy of their own generated outputs. 

\subsection{Backdoor Attack Detection}
Backdoor attacks in LLMs involve inserting malicious triggers during model training, allowing adversaries to manipulate the model's behavior when these triggers are encountered at inference time. To address this issue, researchers have developed various detection and defense mechanisms aimed at identifying and neutralizing backdoor triggers.

The ONION algorithm~\cite{qi2020onion} detects backdoor triggers by calculating the impact of different tokens on a sample's perplexity. BDDR~\cite{SHAO2021102433} identifies potential triggers by analyzing the effect of word removal on model confidence and prevents backdoor activation by removing the trigger and reconstructing the original sample. RAP~\cite{yang2021rap} employs word-based robustness-aware perturbations to compute the confidence difference between original and perturbed samples under the target label, effectively detecting poisoned samples.

Moreover, recent studies have also made efforts to detect backdoor attacks. BDMMT~\cite{10471589} proposes a defense method based on deep model mutation testing, where random mutations are applied to pre-trained language models, and backdoor samples exhibit increased robustness under these mutations, enabling detection. CLEANGEN~\cite{li2024cleangenmitigatingbackdoorattacks} detects and replaces suspicious tokens favored by attackers, while preserving model utility when processing benign user queries. Chain-of-Scrutiny (CoS)~\cite{li2024chainofscrutinydetectingbackdoorattacks} guides the model to generate detailed reasoning steps and examines the consistency between the reasoning process and the final output to identify potential backdoor attacks.

\zsd{\subsection{Broader Security Defenses}
In broader security contexts, such as Industrial Control Systems (ICS) and Federated Learning, advanced defense mechanisms have provided valuable methodological precedents. Notable works, including MoCC-BD-FID~\cite{DBLP:journals/tifs/ZengSLGW25} and MoNAS-IDSAA~\cite{https://doi.org/10.1049/csy2.12117}, effectively utilize multi-objective clustering and neural architecture search to identify backdoors and intrusions in continuous signal systems. While these approaches validate the utility of clustering and architectural optimization for system security, LLM anomaly detection entails unique challenges. Distinct from the numerical time-series data characteristic of ICS, LLMs operate on high-dimensional, discrete textual representations. Consequently, effective defense in this domain necessitates forensic methods specifically tailored to the semantic activation patterns inherent to the Transformer architecture.
}
\section{Conclusion}
\label{sec:conclusion}
In this study, we developed a comprehensive real-time detection framework tailored to address multiple abnormal behaviors in large language models, including hallucinations, jailbreak attacks, and backdoor threats. By analyzing neural activation patterns, particularly in critical layers, our approach captures distinguishing features between normal and abnormal outputs. Specifically, our framework leverages two sets of features including \emph{\textbf{NAS}} and \emph{\textbf{ANE}}, enabling a lightweight classifier to achieve real-time anomaly detection without compromising accuracy. Extensive evaluations across diverse tasks demonstrate the framework’s robustness and generalizability, achieving average accuracies of 97.43\%, 82.69\%, and 91.96\% for detecting jailbreak, hallucination, and backdoor threats, respectively. The framework only incurs minimal computational overhead, enabling real-time detection in deployment. Our work marks a step towards advancing LLM security, offering a scalable and efficient solution for detecting abnormal behavior in constantly-evolving AI applications.

\bibliographystyle{IEEEtran}
\bibliography{ref}

@article{farquhar2024detecting,
  title={Detecting hallucinations in large language models using semantic entropy},
  author={Farquhar, Sebastian and Kossen, Jannik and Kuhn, Lorenz and Gal, Yarin},
  journal={Nature},
  volume={630},
  number={8017},
  pages={625--630},
  year={2024},
  publisher={Nature Publishing Group UK London}
}

@inproceedings{SvyatkovskiyDFS20,
  author       = {Alexey Svyatkovskiy and
                  Shao Kun Deng and
                  Shengyu Fu and
                  Neel Sundaresan},
  editor       = {Prem Devanbu and
                  Myra B. Cohen and
                  Thomas Zimmermann},
  title        = {IntelliCode compose: code generation using transformer},
  booktitle    = {{ESEC/FSE} '20: 28th {ACM} Joint European Software Engineering Conference
                  and Symposium on the Foundations of Software Engineering, Virtual
                  Event, USA, November 8-13, 2020},
  pages        = {1433--1443},
  publisher    = {{ACM}},
  year         = {2020},
  url          = {https://doi.org/10.1145/3368089.3417058},
  doi          = {10.1145/3368089.3417058}
}

@misc{jain2023baseline,
      title={Baseline Defenses for Adversarial Attacks Against Aligned Language Models}, 
      author={Neel Jain and Avi Schwarzschild and Yuxin Wen and Gowthami Somepalli and John Kirchenbauer and Ping-yeh Chiang and Micah Goldblum and Aniruddha Saha and Jonas Geiping and Tom Goldstein},
      year={2023},
      eprint={2309.00614},
      archivePrefix={arXiv},
      primaryClass={cs.LG}
}

@inproceedings{xie-etal-2024-gradsafe,
    title = "{G}rad{S}afe: Detecting Jailbreak Prompts for {LLM}s via Safety-Critical Gradient Analysis",
    author = "Xie, Yueqi  and
      Fang, Minghong  and
      Pi, Renjie  and
      Gong, Neil",
    editor = "Ku, Lun-Wei  and
      Martins, Andre  and
      Srikumar, Vivek",
    booktitle = "Proceedings of the 62nd Annual Meeting of the Association for Computational Linguistics (Volume 1: Long Papers)",
    month = aug,
    year = "2024",
    address = "Bangkok, Thailand",
    publisher = "Association for Computational Linguistics",
    url = "https://aclanthology.org/2024.acl-long.30",
    doi = "10.18653/v1/2024.acl-long.30",
    pages = "507--518",
}

@misc{hu2024gradientcuffdetectingjailbreak,
      title={Gradient Cuff: Detecting Jailbreak Attacks on Large Language Models by Exploring Refusal Loss Landscapes}, 
      author={Xiaomeng Hu and Pin-Yu Chen and Tsung-Yi Ho},
      year={2024},
      eprint={2403.00867},
      archivePrefix={arXiv},
      primaryClass={cs.CR},
      url={https://arxiv.org/abs/2403.00867}, 
}

@misc{inan2023llamaguardllmbasedinputoutput,
      title={Llama Guard: LLM-based Input-Output Safeguard for Human-AI Conversations}, 
      author={Hakan Inan and Kartikeya Upasani and Jianfeng Chi and Rashi Rungta and Krithika Iyer and Yuning Mao and Michael Tontchev and Qing Hu and Brian Fuller and Davide Testuggine and Madian Khabsa},
      year={2023},
      eprint={2312.06674},
      archivePrefix={arXiv},
      primaryClass={cs.CL},
      url={https://arxiv.org/abs/2312.06674}, 
}

@misc{azaria2023internalstatellmknows,
      title={The Internal State of an LLM Knows When It's Lying}, 
      author={Amos Azaria and Tom Mitchell},
      year={2023},
      eprint={2304.13734},
      archivePrefix={arXiv},
      primaryClass={cs.CL},
      url={https://arxiv.org/abs/2304.13734}, 
}

@misc{ravi2024lynxopensourcehallucination,
      title={Lynx: An Open Source Hallucination Evaluation Model}, 
      author={Selvan Sunitha Ravi and Bartosz Mielczarek and Anand Kannappan and Douwe Kiela and Rebecca Qian},
      year={2024},
      eprint={2407.08488},
      archivePrefix={arXiv},
      primaryClass={cs.AI},
      url={https://arxiv.org/abs/2407.08488}, 
}

@misc{duan2024llmsknowhallucinationempirical,
      title={Do LLMs Know about Hallucination? An Empirical Investigation of LLM's Hidden States}, 
      author={Hanyu Duan and Yi Yang and Kar Yan Tam},
      year={2024},
      eprint={2402.09733},
      archivePrefix={arXiv},
      primaryClass={cs.CL},
      url={https://arxiv.org/abs/2402.09733}, 
}

@misc{chen2024incontextsharpnessalertsinner,
      title={In-Context Sharpness as Alerts: An Inner Representation Perspective for Hallucination Mitigation}, 
      author={Shiqi Chen and Miao Xiong and Junteng Liu and Zhengxuan Wu and Teng Xiao and Siyang Gao and Junxian He},
      year={2024},
      eprint={2403.01548},
      archivePrefix={arXiv},
      primaryClass={cs.CL},
      url={https://arxiv.org/abs/2403.01548}, 
}

@misc{ji2024llminternalstatesreveal,
      title={LLM Internal States Reveal Hallucination Risk Faced With a Query}, 
      author={Ziwei Ji and Delong Chen and Etsuko Ishii and Samuel Cahyawijaya and Yejin Bang and Bryan Wilie and Pascale Fung},
      year={2024},
      eprint={2407.03282},
      archivePrefix={arXiv},
      primaryClass={cs.CL},
      url={https://arxiv.org/abs/2407.03282}, 
}

@inproceedings{chen-mueller-2024-quantifying,
    title = "Quantifying Uncertainty in Answers from any Language Model and Enhancing their Trustworthiness",
    author = "Chen, Jiuhai  and
      Mueller, Jonas",
    editor = "Ku, Lun-Wei  and
      Martins, Andre  and
      Srikumar, Vivek",
    booktitle = "Proceedings of the 62nd Annual Meeting of the Association for Computational Linguistics (Volume 1: Long Papers)",
    month = aug,
    year = "2024",
    address = "Bangkok, Thailand",
    publisher = "Association for Computational Linguistics",
    url = "https://aclanthology.org/2024.acl-long.283",
    doi = "10.18653/v1/2024.acl-long.283",
    pages = "5186--5200",
}

@inproceedings{zhang-etal-2024-self,
    title = "Self-Alignment for Factuality: Mitigating Hallucinations in {LLM}s via Self-Evaluation",
    author = "Zhang, Xiaoying  and
      Peng, Baolin  and
      Tian, Ye  and
      Zhou, Jingyan  and
      Jin, Lifeng  and
      Song, Linfeng  and
      Mi, Haitao  and
      Meng, Helen",
    editor = "Ku, Lun-Wei  and
      Martins, Andre  and
      Srikumar, Vivek",
    booktitle = "Proceedings of the 62nd Annual Meeting of the Association for Computational Linguistics (Volume 1: Long Papers)",
    month = aug,
    year = "2024",
    address = "Bangkok, Thailand",
    publisher = "Association for Computational Linguistics",
    url = "https://aclanthology.org/2024.acl-long.107",
    doi = "10.18653/v1/2024.acl-long.107",
    pages = "1946--1965",
}

@article{qi2020onion,
  title={Onion: A simple and effective defense against textual backdoor attacks},
  author={Qi, Fanchao and Chen, Yangyi and Li, Mukai and Yao, Yuan and Liu, Zhiyuan and Sun, Maosong},
  journal={arXiv preprint arXiv:2011.10369},
  year={2020}
}

@article{yang2021rap,
  title={RAP: Robustness-Aware Perturbations for Defending against Backdoor Attacks on NLP Models},
  author={Yang, Wenkai and Lin, Yankai and Li, Peng and Zhou, Jie and Sun, Xu},
  journal={arXiv preprint arXiv:2110.07831},
  year={2021}
}

@article{SHAO2021102433,
title = {BDDR: An Effective Defense Against Textual Backdoor Attacks},
journal = {Computers \& Security},
volume = {110},
pages = {102433},
year = {2021},
issn = {0167-4048},
doi = {https://doi.org/10.1016/j.cose.2021.102433},
url = {https://www.sciencedirect.com/science/article/pii/S0167404821002571},
author = {Kun Shao and Junan Yang and Yang Ai and Hui Liu and Yu Zhang}
}

@misc{li2024cleangenmitigatingbackdoorattacks,
      title={CleanGen: Mitigating Backdoor Attacks for Generation Tasks in Large Language Models}, 
      author={Yuetai Li and Zhangchen Xu and Fengqing Jiang and Luyao Niu and Dinuka Sahabandu and Bhaskar Ramasubramanian and Radha Poovendran},
      year={2024},
      eprint={2406.12257},
      archivePrefix={arXiv},
      primaryClass={cs.AI},
      url={https://arxiv.org/abs/2406.12257}, 
}

@misc{li2024chainofscrutinydetectingbackdoorattacks,
      title={Chain-of-Scrutiny: Detecting Backdoor Attacks for Large Language Models}, 
      author={Xi Li and Yusen Zhang and Renze Lou and Chen Wu and Jiaqi Wang},
      year={2024},
      eprint={2406.05948},
      archivePrefix={arXiv},
      primaryClass={cs.CR},
      url={https://arxiv.org/abs/2406.05948}, 
}

@ARTICLE{10471589,
  author={Wei, Jiali and Fan, Ming and Jiao, Wenjing and Jin, Wuxia and Liu, Ting},
  journal={IEEE Transactions on Information Forensics and Security}, 
  title={BDMMT: Backdoor Sample Detection for Language Models Through Model Mutation Testing}, 
  year={2024},
  volume={19},
  number={},
  pages={4285-4300},
  doi={10.1109/TIFS.2024.3376968}}

@misc{gu2019badnetsidentifyingvulnerabilitiesmachine,
      title={BadNets: Identifying Vulnerabilities in the Machine Learning Model Supply Chain}, 
      author={Tianyu Gu and Brendan Dolan-Gavitt and Siddharth Garg},
      year={2019},
      eprint={1708.06733},
      archivePrefix={arXiv},
      primaryClass={cs.CR},
      url={https://arxiv.org/abs/1708.06733}, 
}

@misc{touvron2023llama,
      title={LLaMA: Open and Efficient Foundation Language Models}, 
      author={Hugo Touvron and Thibaut Lavril and Gautier Izacard and Xavier Martinet and Marie-Anne Lachaux and Timothée Lacroix and Baptiste Rozière and Naman Goyal and Eric Hambro and Faisal Azhar and Aurelien Rodriguez and Armand Joulin and Edouard Grave and Guillaume Lample},
      year={2023},
      eprint={2302.13971},
      archivePrefix={arXiv},
      primaryClass={cs.CL}
}

@misc{Llama3,
  author       = {AI@Meta},
  title        = {Llama 3 model card},
  year         = {2024},
  url          = {https://github.com/meta-llama/llama3/blob/main/MODEL_CARD.md},
  note         = {Accessed: 2025-1-7}
}

@misc{gemmateam2024gemmaopenmodelsbased,
      title={Gemma: Open Models Based on Gemini Research and Technology}, 
      author={Gemma Team and Thomas Mesnard and Cassidy Hardin and Robert Dadashi and Surya Bhupatiraju and Shreya Pathak and Laurent Sifre and Morgane Rivière and Mihir Sanjay Kale and Juliette Love and et al.},
      year={2024},
      eprint={2403.08295},
      archivePrefix={arXiv},
      primaryClass={cs.CL},
      url={https://arxiv.org/abs/2403.08295}, 
}

@misc{peng2023instructiontuninggpt4,
      title={Instruction Tuning with GPT-4}, 
      author={Baolin Peng and Chunyuan Li and Pengcheng He and Michel Galley and Jianfeng Gao},
      year={2023},
      eprint={2304.03277},
      archivePrefix={arXiv},
      primaryClass={cs.CL},
      url={https://arxiv.org/abs/2304.03277}, 
}

@misc{luo2024jailbreakv28kbenchmarkassessingrobustness,
      title={JailBreakV-28K: A Benchmark for Assessing the Robustness of MultiModal Large Language Models against Jailbreak Attacks}, 
      author={Weidi Luo and Siyuan Ma and Xiaogeng Liu and Xiaoyu Guo and Chaowei Xiao},
      year={2024},
      eprint={2404.03027},
      archivePrefix={arXiv},
      primaryClass={cs.CR},
      url={https://arxiv.org/abs/2404.03027}, 
}

@misc{zou2023universal,
      title={Universal and Transferable Adversarial Attacks on Aligned Language Models}, 
      author={Andy Zou and Zifan Wang and J. Zico Kolter and Matt Fredrikson},
      year={2023},
      eprint={2307.15043},
      archivePrefix={arXiv},
      primaryClass={cs.CL}
}

@article{guo2024cold,
  title={Cold-attack: Jailbreaking llms with stealthiness and controllability},
  author={Guo, Xingang and Yu, Fangxu and Zhang, Huan and Qin, Lianhui and Hu, Bin},
  journal={arXiv preprint arXiv:2402.08679},
  year={2024}
}

@article{andriushchenko2024jailbreaking,
      title={Jailbreaking Leading Safety-Aligned LLMs with Simple Adaptive Attacks}, 
      author={Andriushchenko, Maksym and Croce, Francesco and Flammarion, Nicolas},
      journal={arXiv preprint arXiv:2404.02151},
      year={2024}
}

@misc{lin2021truthfulqa,
    title={TruthfulQA: Measuring How Models Mimic Human Falsehoods},
    author={Stephanie Lin and Jacob Hilton and Owain Evans},
    year={2021},
    eprint={2109.07958},
    archivePrefix={arXiv},
    primaryClass={cs.CL}
}

@misc{HaluEval,
  author = {Junyi Li and Xiaoxue Cheng and Wayne Xin Zhao and Jian-Yun Nie and Ji-Rong Wen },
  title = {HaluEval: A Large-Scale Hallucination Evaluation Benchmark for Large Language Models},
  year = {2023},
  journal={arXiv preprint arXiv:2305.11747},
  url={https://arxiv.org/abs/2305.11747}
}

@article{10.1145/3689776,
author = {Li, Ningke and Li, Yuekang and Liu, Yi and Shi, Ling and Wang, Kailong and Wang, Haoyu},
title = {Drowzee: Metamorphic Testing for Fact-Conflicting Hallucination Detection in Large Language Models},
year = {2024},
issue_date = {October 2024},
publisher = {Association for Computing Machinery},
address = {New York, NY, USA},
volume = {8},
number = {OOPSLA2},
url = {https://doi.org/10.1145/3689776},
doi = {10.1145/3689776},
journal = {Proc. ACM Program. Lang.},
month = oct,
articleno = {336},
numpages = {30}
}

@inproceedings{yan-etal-2024-backdooring,
    title = "Backdooring Instruction-Tuned Large Language Models with Virtual Prompt Injection",
    author = "Yan, Jun  and
      Yadav, Vikas  and
      Li, Shiyang  and
      Chen, Lichang  and
      Tang, Zheng  and
      Wang, Hai  and
      Srinivasan, Vijay  and
      Ren, Xiang  and
      Jin, Hongxia",
    editor = "Duh, Kevin  and
      Gomez, Helena  and
      Bethard, Steven",
    booktitle = "Proceedings of the 2024 Conference of the North American Chapter of the Association for Computational Linguistics: Human Language Technologies (Volume 1: Long Papers)",
    month = jun,
    year = "2024",
    address = "Mexico City, Mexico",
    publisher = "Association for Computational Linguistics",
    url = "https://aclanthology.org/2024.naacl-long.337",
    doi = "10.18653/v1/2024.naacl-long.337",
    pages = "6065--6086",
}

@misc{li2024backdoorllmcomprehensivebenchmarkbackdoor,
      title={BackdoorLLM: A Comprehensive Benchmark for Backdoor Attacks on Large Language Models}, 
      author={Yige Li and Hanxun Huang and Yunhan Zhao and Xingjun Ma and Jun Sun},
      year={2024},
      eprint={2408.12798},
      archivePrefix={arXiv},
      primaryClass={cs.AI},
      url={https://arxiv.org/abs/2408.12798}, 
}

@inproceedings{10.1145/3132747.3132785,
author = {Pei, Kexin and Cao, Yinzhi and Yang, Junfeng and Jana, Suman},
title = {DeepXplore: Automated Whitebox Testing of Deep Learning Systems},
year = {2017},
isbn = {9781450350853},
publisher = {Association for Computing Machinery},
address = {New York, NY, USA},
url = {https://doi.org/10.1145/3132747.3132785},
doi = {10.1145/3132747.3132785},
booktitle = {Proceedings of the 26th Symposium on Operating Systems Principles},
pages = {1–18},
numpages = {18},
location = {Shanghai, China},
series = {SOSP '17}
}

@inproceedings{10.1145/3238147.3238202,
author = {Ma, Lei and Juefei-Xu, Felix and Zhang, Fuyuan and Sun, Jiyuan and Xue, Minhui and Li, Bo and Chen, Chunyang and Su, Ting and Li, Li and Liu, Yang and Zhao, Jianjun and Wang, Yadong},
title = {DeepGauge: multi-granularity testing criteria for deep learning systems},
year = {2018},
isbn = {9781450359375},
publisher = {Association for Computing Machinery},
address = {New York, NY, USA},
url = {https://doi.org/10.1145/3238147.3238202},
doi = {10.1145/3238147.3238202},
booktitle = {Proceedings of the 33rd ACM/IEEE International Conference on Automated Software Engineering},
pages = {120–131},
numpages = {12},
location = {Montpellier, France},
series = {ASE '18}
}

@INPROCEEDINGS{7081877,
  author={Kochhar, Pavneet Singh and Thung, Ferdian and Lo, David},
  booktitle={2015 IEEE 22nd International Conference on Software Analysis, Evolution, and Reengineering (SANER)}, 
  title={Code coverage and test suite effectiveness: Empirical study with real bugs in large systems}, 
  year={2015},
  volume={},
  number={},
  pages={560-564},
  doi={10.1109/SANER.2015.7081877}}

@inproceedings{DBLP:conf/wsdm/GoyalRRYZCNW24,
  author       = {Sagar Goyal and
                  Eti Rastogi and
                  Sree Prasanna Rajagopal and
                  Dong Yuan and
                  Fen Zhao and
                  Jai Chintagunta and
                  Gautam Naik and
                  Jeff Ward},
  editor       = {Luz Angelica Caudillo{-}Mata and
                  Silvio Lattanzi and
                  Andr{\'{e}}s Mu{\~{n}}oz Medina and
                  Leman Akoglu and
                  Aristides Gionis and
                  Sergei Vassilvitskii},
  title        = {HealAI: {A} Healthcare {LLM} for Effective Medical Documentation},
  booktitle    = {Proceedings of the 17th {ACM} International Conference on Web Search
                  and Data Mining, {WSDM} 2024, Merida, Mexico, March 4-8, 2024},
  pages        = {1167--1168},
  publisher    = {{ACM}},
  year         = {2024},
  url          = {https://doi.org/10.1145/3616855.3635739},
  doi          = {10.1145/3616855.3635739},
  bibsource    = {dblp computer science bibliography, https://dblp.org}
}

@inproceedings{DBLP:conf/icaif/LiWDC23,
  author       = {Yinheng Li and
                  Shaofei Wang and
                  Han Ding and
                  Hang Chen},
  title        = {Large Language Models in Finance: {A} Survey},
  booktitle    = {4th {ACM} International Conference on {AI} in Finance, {ICAIF} 2023,
                  Brooklyn, NY, USA, November 27-29, 2023},
  pages        = {374--382},
  publisher    = {{ACM}},
  year         = {2023},
  url          = {https://doi.org/10.1145/3604237.3626869},
  doi          = {10.1145/3604237.3626869},
  bibsource    = {dblp computer science bibliography, https://dblp.org}
}

@article{DBLP:journals/corr/abs-2405-03644,
  author       = {Jie Zhang and
                  Haoyu Bu and
                  Hui Wen and
                  Yu Chen and
                  Lun Li and
                  Hongsong Zhu},
  title        = {When LLMs Meet Cybersecurity: {A} Systematic Literature Review},
  journal      = {CoRR},
  volume       = {abs/2405.03644},
  year         = {2024},
  url          = {https://doi.org/10.48550/arXiv.2405.03644},
  doi          = {10.48550/ARXIV.2405.03644},
  eprinttype    = {arXiv},
  eprint       = {2405.03644},
  bibsource    = {dblp computer science bibliography, https://dblp.org}
}

@article{10.1145/3703155,
author = {Huang, Lei and Yu, Weijiang and Ma, Weitao and Zhong, Weihong and Feng, Zhangyin and Wang, Haotian and Chen, Qianglong and Peng, Weihua and Feng, Xiaocheng and Qin, Bing and Liu, Ting},
title = {A Survey on Hallucination in Large Language Models: Principles, Taxonomy, Challenges, and Open Questions},
year = {2024},
publisher = {Association for Computing Machinery},
address = {New York, NY, USA},
issn = {1046-8188},
url = {https://doi.org/10.1145/3703155},
doi = {10.1145/3703155},
note = {Just Accepted},
journal = {ACM Trans. Inf. Syst.},
month = nov
}

@inproceedings{DBLP:conf/acl/XuLDLP24,
  author       = {Zihao Xu and
                  Yi Liu and
                  Gelei Deng and
                  Yuekang Li and
                  Stjepan Picek},
  editor       = {Lun{-}Wei Ku and
                  Andre Martins and
                  Vivek Srikumar},
  title        = {A Comprehensive Study of Jailbreak Attack versus Defense for Large
                  Language Models},
  booktitle    = {Findings of the Association for Computational Linguistics, {ACL} 2024,
                  Bangkok, Thailand and virtual meeting, August 11-16, 2024},
  pages        = {7432--7449},
  publisher    = {Association for Computational Linguistics},
  year         = {2024},
  url          = {https://doi.org/10.18653/v1/2024.findings-acl.443},
  doi          = {10.18653/V1/2024.FINDINGS-ACL.443},
  bibsource    = {dblp computer science bibliography, https://dblp.org}
}

@inproceedings{DBLP:conf/ndss/DengLLWZLW0L24,
  author       = {Gelei Deng and
                  Yi Liu and
                  Yuekang Li and
                  Kailong Wang and
                  Ying Zhang and
                  Zefeng Li and
                  Haoyu Wang and
                  Tianwei Zhang and
                  Yang Liu},
  title        = {{MASTERKEY:} Automated Jailbreaking of Large Language Model Chatbots},
  booktitle    = {31st Annual Network and Distributed System Security Symposium, {NDSS}
                  2024, San Diego, California, USA, February 26 - March 1, 2024},
  publisher    = {The Internet Society},
  year         = {2024},
  url          = {https://www.ndss-symposium.org/ndss-paper/masterkey-automated-jailbreaking-of-large-language-model-chatbots/},
  bibsource    = {dblp computer science bibliography, https://dblp.org}
}

@misc{zhou2025understandingeffectivenesscoveragecriteria,
      title={Understanding the Effectiveness of Coverage Criteria for Large Language Models: A Special Angle from Jailbreak Attacks}, 
      author={Shide Zhou and Tianlin Li and Kailong Wang and Yihao Huang and Ling Shi and Yang Liu and Haoyu Wang},
      year={2025},
      eprint={2408.15207},
      archivePrefix={arXiv},
      primaryClass={cs.SE},
      url={https://arxiv.org/abs/2408.15207}, 
}

@article{DBLP:journals/tifs/ZengSLGW25,
  author       = {Guo{-}Qiang Zeng and
                  Jun{-}Min Shao and
                  Kang{-}Di Lu and
                  Guang{-}Gang Geng and
                  Jian Weng},
  title        = {MoCC-BD-FID: Multi-Objective Clustering Combination-Based Backdoor
                  Defense for Federated Intrusion Detection of Industrial Control Systems},
  journal      = {{IEEE} Trans. Inf. Forensics Secur.},
  volume       = {20},
  pages        = {6868--6883},
  year         = {2025},
  url          = {https://doi.org/10.1109/TIFS.2025.3586479},
  doi          = {10.1109/TIFS.2025.3586479},
  bibsource    = {dblp computer science bibliography, https://dblp.org}
}

@article{https://doi.org/10.1049/csy2.12117,
author = {Zeng, Guo-Qiang and Shao, Jun-Min and Lu, Kang-Di and Geng, Guang-Gang and Weng, Jian},
title = {Automated federated learning-based adversarial attack and defence in industrial control systems},
journal = {IET Cyber-Systems and Robotics},
volume = {6},
number = {2},
pages = {e12117},
doi = {https://doi.org/10.1049/csy2.12117},
url = {https://ietresearch.onlinelibrary.wiley.com/doi/abs/10.1049/csy2.12117},
eprint = {https://ietresearch.onlinelibrary.wiley.com/doi/pdf/10.1049/csy2.12117},
year = {2024}
}

@INPROCEEDINGS{10172609,
  author={Ji, Zhenlan and Ma, Pingchuan and Yuan, Yuanyuan and Wang, Shuai},
  booktitle={2023 IEEE/ACM 45th International Conference on Software Engineering (ICSE)}, 
  title={CC: Causality-Aware Coverage Criterion for Deep Neural Networks}, 
  year={2023},
  volume={},
  number={},
  pages={1788-1800},
  doi={10.1109/ICSE48619.2023.00153}}

@INPROCEEDINGS{10172683,
  author={Yuan, Yuanyuan and Pang, Qi and Wang, Shuai},
  booktitle={2023 IEEE/ACM 45th International Conference on Software Engineering (ICSE)}, 
  title={Revisiting Neuron Coverage for DNN Testing: A Layer-Wise and Distribution-Aware Criterion}, 
  year={2023},
  volume={},
  number={},
  pages={1200-1212},
  doi={10.1109/ICSE48619.2023.00107}}

@misc{wildteaming2024,
      title={WildTeaming at Scale: From In-the-Wild Jailbreaks to (Adversarially) Safer Language Models}, 
      author={Liwei Jiang and Kavel Rao and Seungju Han and Allyson Ettinger and Faeze Brahman and Sachin Kumar and Niloofar Mireshghallah and Ximing Lu and Maarten Sap and Yejin Choi and Nouha Dziri},
      year={2024},
      eprint={2406.18510},
      archivePrefix={arXiv},
      primaryClass={cs.CL},
      url={https://arxiv.org/abs/2406.18510}, 
}

@inproceedings{SciQ,
    title={Crowdsourcing Multiple Choice Science Questions},
    author={Johannes Welbl, Nelson F. Liu, Matt Gardner},
    year={2017},
    journal={arXiv:1707.06209v1}
}

@misc{abnordetector_site,
  author = {{LLM-Abnormal-Detection}},
  title = {Exposing the Ghost in the Transformer: Abnormal Detection for Large Language Models via Hidden State Forensics},
  year = {2026},
  url = {https://sites.google.com/view/llm-abnormal-detection},
  note = {Accessed: 2026-1-10}
}

@inproceedings{DBLP:conf/ccs/ShenC0SZ24,
  author       = {Xinyue Shen and
                  Zeyuan Chen and
                  Michael Backes and
                  Yun Shen and
                  Yang Zhang},
  editor       = {Bo Luo and
                  Xiaojing Liao and
                  Jun Xu and
                  Engin Kirda and
                  David Lie},
  title        = {"Do Anything Now": Characterizing and Evaluating In-The-Wild Jailbreak
                  Prompts on Large Language Models},
  booktitle    = {Proceedings of the 2024 on {ACM} {SIGSAC} Conference on Computer and
                  Communications Security, {CCS} 2024, Salt Lake City, UT, USA, October
                  14-18, 2024},
  pages        = {1671--1685},
  publisher    = {{ACM}},
  year         = {2024},
  url          = {https://doi.org/10.1145/3658644.3670388},
  doi          = {10.1145/3658644.3670388},
  bibsource    = {dblp computer science bibliography, https://dblp.org}
}

@article{DBLP:journals/corr/abs-2506-06518,
  author       = {Neil Fendley and
                  Edward W. Staley and
                  Joshua Carney and
                  William Redman and
                  Marie Chau and
                  Nathan Drenkow},
  title        = {A Systematic Review of Poisoning Attacks Against Large Language Models},
  journal      = {CoRR},
  volume       = {abs/2506.06518},
  year         = {2025},
  url          = {https://doi.org/10.48550/arXiv.2506.06518},
  doi          = {10.48550/ARXIV.2506.06518},
  eprinttype    = {arXiv},
  eprint       = {2506.06518},
  bibsource    = {dblp computer science bibliography, https://dblp.org}
}

@article{DBLP:journals/corr/abs-2401-01301,
  author       = {Matthew Dahl and
                  Varun Magesh and
                  Mirac Suzgun and
                  Daniel E. Ho},
  title        = {Large Legal Fictions: Profiling Legal Hallucinations in Large Language
                  Models},
  journal      = {CoRR},
  volume       = {abs/2401.01301},
  year         = {2024},
  url          = {https://doi.org/10.48550/arXiv.2401.01301},
  doi          = {10.48550/ARXIV.2401.01301},
  eprinttype    = {arXiv},
  eprint       = {2401.01301},
  bibsource    = {dblp computer science bibliography, https://dblp.org}
}

@article{DBLP:journals/corr/abs-2505-24830,
  author       = {Juraj Vladika and
                  Annika Domres and
                  Mai Nguyen and
                  Rebecca Moser and
                  Jana Nano and
                  Felix Busch and
                  Lisa C. Adams and
                  Keno K. Bressem and
                  Denise Bernhardt and
                  Stephanie E. Combs and
                  Kai J. Borm and
                  Florian Matthes and
                  Jan C. Peeken},
  title        = {Improving Reliability and Explainability of Medical Question Answering
                  through Atomic Fact Checking in Retrieval-Augmented LLMs},
  journal      = {CoRR},
  volume       = {abs/2505.24830},
  year         = {2025},
  url          = {https://doi.org/10.48550/arXiv.2505.24830},
  doi          = {10.48550/ARXIV.2505.24830},
  eprinttype    = {arXiv},
  eprint       = {2505.24830},
  bibsource    = {dblp computer science bibliography, https://dblp.org}
}

\end{document}